\documentclass[twocolumn]{aastex63}

\usepackage{amsmath}
\usepackage{hyperref}
\usepackage{graphicx}
\usepackage[maxfloats=40]{morefloats}
\usepackage{multirow}
\usepackage[para,online,flushleft]{threeparttable}

\definecolor{green}{rgb}{0.13,0.55,0.13}
\newcommand{\add}[1]{{{#1}}}
\newcommand{\addb}[1]{{{#1}}}

\newcommand{\lya}{Ly$\alpha$}
\newcommand{\SBunits}{{\rm erg\,s^{-1} cm^{-2} arcsec^{-2}}}
\newcommand{\SBunitsA}{{\rm erg\,s^{-1} cm^{-2}\textup{\AA}^{-1} arcsec^{-2}}}
\newcommand{\hinvMpc}{h^{-1}{\rm Mpc}}
\newcommand{\REW}{{\rm REW}}
\newcommand{\UV}{{\rm UV}}

\newcommand{\valSBzero}{3.49_{-2.02}^{+2.27}}
\newcommand{\valmulya}{1.13^{+0.57}_{-0.53}}
\newcommand{\valbetalya}{0.07^{+1.65}_{-0.73}}
\newcommand{\valrholya}{6.6_{-3.1}^{+3.3}}
\newcommand{\valnq}{1.34\times10^{-4}}
\newcommand{\vallogLq}{45.12^{+0.18}_{-0.27}}
\newcommand{\valSFRD}{0.06\pm0.03}

\received{xx, xx}
\revised{xx, xx}
\accepted{xx}
\submitjournal{ApJS}

\shorttitle{Probing the Iceberg of Ly$\alpha$ Emission through Intensity Mapping}
\shortauthors{X. Lin et al.}

\graphicspath{{./}{figures/}}

\begin{document}

\title{Probing the Diffuse Lyman-alpha Emission on Cosmological Scales: Ly$\alpha$ Emission Intensity Mapping
Using the Complete SDSS-IV eBOSS Survey}

\author[0000-0001-6052-4234]{Xiaojing Lin}
\affiliation{Department of Astronomy, 
Tsinghua University, Beijing 100084, China
\email{linxj21@mails.tsinghua.edu.cn, zcai@mail.tsinghua.edu.cn}
}

\author[0000-0003-1887-6732]{Zheng Zheng}
\affiliation{Department of Physics and Astronomy, University of Utah, UT 84112, USA\email{zhengzheng@astro.utah.edu}}

\author[0000-0001-8467-6478]{Zheng Cai}
\affiliation{Department of Astronomy, 
Tsinghua University, Beijing 100084, China\email{zcai@mail.tsinghua.edu.cn}}

\begin{abstract}
Based on the Sloan Digital Sky Survey Data Release 16, we have detected the large-scale structure of Ly$\alpha$ emission in the Universe at redshifts $z = 2$--3.5 by cross-correlating quasar positions and  \lya\ emission imprinted in the residual spectra of luminous red galaxies. We apply an analytical model to fit the corresponding \lya\ surface brightness profile and multipoles of the redshift-space quasar-\lya\ emission cross-correlation function. 
The model suggests an average cosmic \lya\ luminosity density of $\valrholya\times 10^{40} {\rm erg\, s^{-1} cMpc^{-3}}$, \add{a $\sim 2\sigma$ detection with a median value about 8--9 times those estimated from deep narrowband surveys of \lya\ emitters at similar redshifts. Although the low signal-to-noise ratio prevents us from a significant detection of the \lya\ forest-\lya\ emission cross-correlation, the measurement is consistent with the prediction of our best-fit model from quasar-\lya\ emission cross-correlation within current uncertainties}. We rule out the scenario that these \lya\ photons mainly originate from quasars. We find that \lya\ emission from star-forming galaxies, including contributions from that concentrated around the galaxy centers and that in the diffuse \lya\ emitting halos, is able to explain the bulk of the the \lya\ luminosity density inferred from our measurements. Ongoing and future surveys can further improve the measurements and advance our understanding of the cosmic \lya\ emission field.

\end{abstract}


\section{Introduction} \label{sec:intro}

The filamentary structure of the cosmic web, which links galaxies to the intergalactic medium (IGM), is predicted to be a rich reservoir of nearly pristine gas \citep[e.g.,][]{Fumagalli2011,Giavalisco2011ApJ}. Reprocessed radiation from quasars or the ultraviolet (UV) background will ionize hydrogen atoms in the circumgalactic medium (CGM) and IGM \citep[e.g.,][]{Gallego,Borisova2016,Niemeyer2022}, and the recombination of the ionized hydrogen will produce fluorescent Ly$\alpha$ emission, especially in the high-redshift Universe \citep[]{Cantalupo2008,Li2021}.  Extended Ly$\alpha$ emission is expected due to the high cross-section of Ly$\alpha$ photons for resonant scatterings by neutral hydrogen \citep{Zheng2011a}.

Direct imaging of the IGM Ly$\alpha$ emission is challenging because of its low surface brightness (SB)\citep[]{Cantalupo2005}.  One solution is to search around local ionized sources, such as luminous quasars, which reside at the densest regions of the cosmic web. The diffuse gas emission can be enhanced by orders of magnitude, leading to the discovery of the enormous Ly$\alpha$ nebulae (ELANe) \citep{cantalupo2014,Hennawi2015,Cai2017,Cai2018,Fab2019}.  These extrema of Ly$\alpha$ nebulosities have \lya\ surface brightness $\geq$ $10^{-17}\SBunits$  and \lya\ luminosity $\geq$ $10^{44} {\rm erg\, s^{-1}}$, with Ly$\alpha$ size greater than 200 kpc. Recently, the progress in wide-field integral field spectrographs extends the detectability of CGM/IGM with low surface brightness. The most advanced facilities, such as the Keck Cosmic Web Imager \citep[KCWI,][]{Morrissey2018} and the MultiUnit Spectroscopic Explorer \citep[MUSE,][]{Bacon2010}, can reach a surface brightness of a few $\times 10^{-19}\SBunits$, making it possible for observational probes into emission from the CGM/IGM in the vicinity of bright sources (KCWI: \citealt{Borisova2016,Fab2016,Cai2019}, etc.; MUSE: \citealt{Wisotzki2018,Bacon2021,Kusakabe2022}, etc.).  Large amounts of individual \lya\ halos around strong Ly$\alpha$ emitters (LAEs), have been detected thanks to these state-of-the-art instruments \citep[e.g.,][]{Wisotzki2016,Leclercq2017}. Moreover, a recent discovery unveiled that star-forming galaxies generally have \lya\ halos by investigating \lya\ emission around UV-selected galaxies  \citep{Kusakabe2022}.

On scales up to several Mpc from the central bright sources, no direct observational evidence for diffuse gas emissions is found so far. The predicted Ly$\alpha$ surface brightness at $z\geq3$ stimulated by the diffuse ionizing background, is on the order of $10^{-20}$ erg s$^{-1}$ cm$^{-2}$ arcsec$^{-2}$ \citep{Gould1996,Cantalupo2005,Kollmeier2010,Witstok2019}. Currently, this goes far beyond the capability of the most advanced instruments on individual detections. The technique of line intensity mapping \citep{Kovetz2017} is expected to exceed current observational limits, by mapping large scale structures with integrated emission from spectral lines originating from galaxies and the diffuse IGM, but without resolving discrete objects. Its application on 21-cm \ion{H}{1} emission has revealed a promising prospect for observing the low-density cosmic web \citep{Masui2013,Anderson2018,Tramonte2019,Tramonte2020}.  

\lya\ line can also be used for intensity mapping. Ly$\alpha$ intensity mapping experiments can provide viable complimentary approaches to testing many theoretical predictions on the diffuse emission from IGM filaments \citep{silva2013,silva2016,Heneka2017,Elias2020}, bringing new insights into the evolution of the Universe independent of cosmological hydrodynamic simulations \citep{Gallego2018,Gallego,Croft2016,Croft2018}. 
\cite{Croft2016} measured the large-scale structure of Ly$\alpha$ emission by the cross-correlation between Ly$\alpha$ surface brightness, extracted from the spectra of luminous red galaxies (LRGs), and quasars in Sloan Digital Sky Survey (SDSS)/Baryon Oscillation Spectroscopic Survey (BOSS). \add{If the \lya\ emission originates from star formation in faint \lya\ emitting galaxies, the star formation rate density (SFRD) inferred from the measurement would be $\sim$30 times higher than those from narrow-band (NB) LAE surveys but comparable to dust corrected UV estimates, \addb{if nearly all the \lya\ photons from
these galaxies escape without dust absorption} \citep{Croft2016}. } 
 They updated their measurements in \cite{Croft2018} using SDSS Data Release 12 (DR12).  After careful examination for possible contaminations and systematics, the corrected cross-correlation is $\sim$50 percent lower than the DR10 result 
of \citet{Croft2016}. They also performed the cross-correlation of \lya\ emission and Ly$\alpha$ forest as complementary evidence, which presents no signal, and claimed that quasars would dominate the Ly$\alpha$ surface brightness within 15 $h^{-1}$ cMpc. 

Inspired by the cross-correlation technique in \cite{Croft2016,Croft2018}, we measure the Ly$\alpha$ surface brightness on scales of several Mpc from quasars using the most up-to-date LRG spectra and quasar catalog in SDSS DR16, much larger samples than those in \cite{Croft2018}. In Section~\ref{sec:data} we introduce the data samples used in this work. We compute the quasar-Ly$\alpha$ emission cross-correlation and obtain the projected surface brightness profile in Section~\ref{sec:xcf}. In Section~\ref{sec:forest correlation}, Ly$\alpha$ forest-Ly$\alpha$ emission cross-correlation is carried out as a complementary measurement. In Section~\ref{sec:discussion} we perform simple analysis on our results and investigate possible Ly$\alpha$ sources for our detected signals. Our methods to remove potential contamination are presented in Appendix \ref{sec:systematics}.  

Throughout this paper, we adopt a spatially flat $\Lambda$ cold dark matter ($\Lambda$CDM) cosmological model \add{according to the Planck 2018 results} \citep{Planck2020}, with $H_0 = 100 h \, {\rm km\, s^{-1} Mpc^{-1}}$ with $h=0.674$, $\Omega_m$ = 0.315, $\Omega_{b}h^2$ = 0.0224, and $\Omega_ch^2$ = 0.120. We use pMpc (physical Mpc) or pkpc (physical kpc) to denote physical distances and cMpc (comoving Mpc) to denote comoving Mpc.

\section{Data Samples}\label{sec:data}

The data used in this study are selected from the final eBOSS data in the SDSS Data Release 16 (DR16; \citealt{SDSS_DR16}), the fourth data release of the fourth phase of the Sloan Digital Sky Survey (SDSS-\uppercase\expandafter{\romannumeral4}), which contains SDSS observations through August 2018.  As the largest volume survey of the Universe up to date, the eBOSS survey is designed to study the expansion and structure growth history of the Universe and constrain the nature of dark energy by spectroscopic observation of galaxies and quasars. The spectrograph for SDSS-\uppercase\expandafter{\romannumeral4} eBOSS covers a wavelength range of 3650-10,400\AA, with a resolution of $\lambda/\Delta\lambda \sim$ 1500 at 3800\AA\ and $\sim$2500 at 9000\AA. There are 1000 fibers per 7-square-degree plate, and each fiber has a diameter of 120 $\mu$m, i.e., 2 arcsec in angle. There are two spectrographs, with each collecting data from 500 fibers, roughly 450 dedicated to science targets and 50 for flux calibration and sky-background subtraction. The eBOSS data from SDSS DR16 also include spectra obtained using the SDSS-I/II spectrographs covering 3800\AA\ to 9100\AA.

In this work, we correlate the residual flux in the galaxy spectra (after subtracting bestfit galaxy spectral templates) with quasars and \lya\ forest to extract information of high-redshift \lya\ emission imprinted in the galaxy fiber spectra. We describe the quasar catalog, the LRG spectra, and the \lya\ forest samples used in this work.

\subsection{Quasar Catalog}
The SDSS DR16 quasar catalog (DR16Q; \citealt{DR16Q}), the largest selection of spectroscopically confirmed quasars to date, contains 750,414 quasars in total, including 225,082 new quasars observed for the first time.  DR16Q includes different redshift estimates generated by different methods, such as the SDSS spectroscopic pipeline, visual inspection and principal component analysis (PCA). It also provides a ``primary'' redshift for each quasar, which is selected from, most preferably, the visual inspection redshift, or, alternatively, the SDSS automated pipeline redshift. In this work we adopt the ``primary'' redshift and apply a redshift restriction of $2.0 \leq z < 3.5$. This redshift cut is also adopted in \citet{Croft2016,Croft2018} due to the spectrograph cutoff for low-redshift \lya\ emission and the limited number of observed quasars at higher redshifts. Further, we exclude quasars with redshift estimates of ``catastrophic failures'', if their PCA-based redshift estimates have a velocity difference of $|\Delta v|>3000$ km s$^{-1}$ from the ``primary'' redshift. We end up with 255,570 quasars in total, with a median redshift of 2.40.

\subsection{LRG Spectra}
For one of the main projects of the SDSS surveys, a large sample of LRGs have been observed spectroscopically to detect the baryon acoustic oscillations
(BAO) feature. BOSS has been conducted during 2009--2014, producing two principal galaxy samples, LOWZ and CMASS \citep{BOSS_galaxy}. The BOSS LOWZ galaxy sample targets the 343,160 low-redshift galaxy population spanning redshifts $0.15 < z < 0.43$, to extend  the SDSS-I/II Cut I LRG sample \citep{Eisenstein2001} by selecting galaxies of dimmer luminosity.  The BOSS CMASS galaxy sample targets 862,735 higher-redshift ($0.43 < z < 0.75$) galaxies. It used similar color-magnitude cuts to those utilized by the Cut-II LRGs from SDSS-I/II and the LRGs in 2SLAQ \citep{Cannon2006}, but with the galaxy selection towards the bluer and fainter galaxies. Operated over 2014--2019, the eBOSS LRG sample \citep{SDSS_DR16} extends the high-redshift tail of the BOSS galaxies, with 298,762 LRGs covering a redshift range of $0.6 < z < 1.0$.

We select 1,389,712 LRG spectra from the combination of BOSS LOWZ sample, BOSS CMASS sample and eBOSS LRG sample. These LRG spectra have been wavelength-calibrated, sky-subtracted, flux-calibrated, and are the co-added ones of at least three individual exposures, with a uniform logarithmic wavelength grid spacing of $\Delta\log_{10}\lambda=10^{-4}$ (about 69 ${\rm km\, s^{-1}}$ per pixel). Each spectrum has an inverse variance per pixel to estimate the uncertainty, which incorporates photon noise, CCD read noise, and sky-subtraction error. Bad pixels are flagged by pixel mask information, and we use \texttt{AND\_MASK} provided by SDSS to rule out bad pixels in all exposures. 

Each LRG spectrum has a best-fitting model spectrum by performing a rest-frame PCA using four eigenspectra as the basis \citep{Bolton}. A set of trial redshifts are explored by shifting the galaxy eigenbasis and modelling their minimum-chi-squared linear combination. A quadratic polynomial is added to fit some low-order calibration uncertainties, such as the Galactic extinction, intrinsic extinction and residual spectrophotometric calibration errors. For each fiber, any objects along the corresponding line of sight that falls within the fiber aperture can have their emission imprinted in the spectrum. For example, the LRG fiber may capture the signal of diffuse \lya\ emission originated from high-redshift galaxies and intergalactic medium, and this is the signal we intend to extract in this work.

In the following analysis we only use the pixels from 3647\AA\ to 5470\AA\ in the observed frame, corresponding to \lya\ emission in the redshift range $2.0 <z < 3.5$.

\subsection{Ly$\alpha$ Forest}
The Ly$\alpha$ forest samples\footnote{\url{https://data.sdss.org/sas/dr16/eboss/lya/Delta_LYA/}} used in this work are selected from the ``Ly$\alpha$ regions'', $\lambda_{\rm RF}\in [1040,1200]$\AA, of 210,005 BOSS/eBOSS quasar spectra ranging from $z=2.1$ to $z=4$ \citep{Bourboux2020}, where $\lambda_{\rm RF}$ represents the wavelength in quasar's rest frame. Broad absorption line quasars (BAL QSOs), bad observations, and spectra whose Ly$\alpha$ regions have less than 50 pixels are all excluded. Then every three original pipeline spectral pixels ($\Delta \log_{10} \lambda\sim 10^{-4}$) are rebinned ($\Delta \log_{10} \lambda\sim 3\times 10^{-4}$) for the purpose of measuring Ly$\alpha$ correlations.

For each spectral region the flux-transmission fields are estimated by the ratio of the observed flux, $f_q$, to the mean expected flux, $\langle F_q \rangle$ \citep{Bourboux2020}:
\begin{equation}
\delta_{f}(\lambda)=\frac{f_{q}}{\langle F_q \rangle}-1.
\end{equation}

The pipeline deals with Ly$\alpha$ forest with identified damped Ly$\alpha$ systems (DLAs) cautiously. Pixels where a DLA reduces the transmission by more than 20\% are masked, and the absorption in the wings is corrected using a Voigt profile following the procedure described in \citet{Noterdaeme2012}. Besides, we also mask $\pm50$\AA\ regions around the DLA positions predicted by \citet{Ho2021}, to ensure that DLA contamination is removed. The number of the remaining Ly$\alpha$ forest pixels is $\sim3.4\times10^7$, with a median redshift of 2.41.

\section{Quasar-Ly$\alpha$ emission Cross-correlation}\label{sec:xcf}

As the SDSS fiber would capture signals from high-redshift background sources, 
the LRG residual spectra, with the bestfit galaxy model spectra subtracted, 
may have Ly$\alpha$ emission from the high-redshift galaxies and IGM/CGM superposed. However, the signals are overwhelmed by noises in most cases. 
Cross-correlating the residual spectrum pixels with quasar 
positions is equivalent to stacking the Ly$\alpha$ signal 
in the quasar neighborhood. Suppressing the noise, the cross-correlation 
technique makes it possible to exceed current observation limits and 
detect diffuse Ly$\alpha$ emission with dimmer luminosities \citep[]{Croft2016,Croft2018}. 

In this section we
perform and analyze the quasar-Ly$\alpha$ cross-correlation using the quasar catalog and LRG spectra mentioned in Section~\ref{sec:data}. In Section~\ref{sec:xiqa} we describe the detailed measurement of the two-dimensional cross-correlation as a function of the separations along and perpendicular to the line-of-sight direction. We measure the corresponding projected surface brightness profile in Section~\ref{sec:proj_SB} and multipoles of the redshift-space two-point correlation function in Section~\ref{sec:2pcf}.

\subsection{Cross-correlation Transverse and Parallel to the Line of Sight}
\label{sec:xiqa}

Firstly we split the LRGs into 885 subsamples based on their angular positions, identified by the HEALPix \citep{Gorski2005} number with N$_{\rm side}$=16, which makes it convenient to search for neighboring quasars within a limited sky region. After obtaining a quasar-LRG spectrum pixel pair with an angular separation of $\theta$, we can compute the line-of-sight separation $r_\|$ and transverse separation $r_{\perp}$ between these two objects:
\begin{eqnarray}
r_\| &=& \left[D_{\rm C}(z_{\rm Ly\alpha})- D_{\rm C}(z_{\rm q}) \right] \cos\frac{\theta}{2}, \\
r_{\perp} &=& \left[D_{\rm M}(z_{\rm Ly\alpha}) + D_{\rm M}(z_{\rm q} \right] \sin{\frac{\theta}{2}},
\end{eqnarray}
where $D_{\rm C}$ is the line-of-sight comoving distance as a function of redshift $z$, $D_{\rm M}$ is the transverse comoving distance as a function of redshift $z$, $z_{\rm q}$ is the quasar redshift and $z_{\rm Ly\alpha}$ is the redshift of Ly$\alpha$ emission converted from the wavelength of the LRG spectrum pixel, i.e., $z_{\rm Ly\alpha}=\lambda/\lambda_{\rm Ly\alpha}-1$ with $\lambda_{\rm Ly\alpha}=1215.67$\AA. 

\begin{figure}
    \centering
    \includegraphics[width=.5\textwidth]{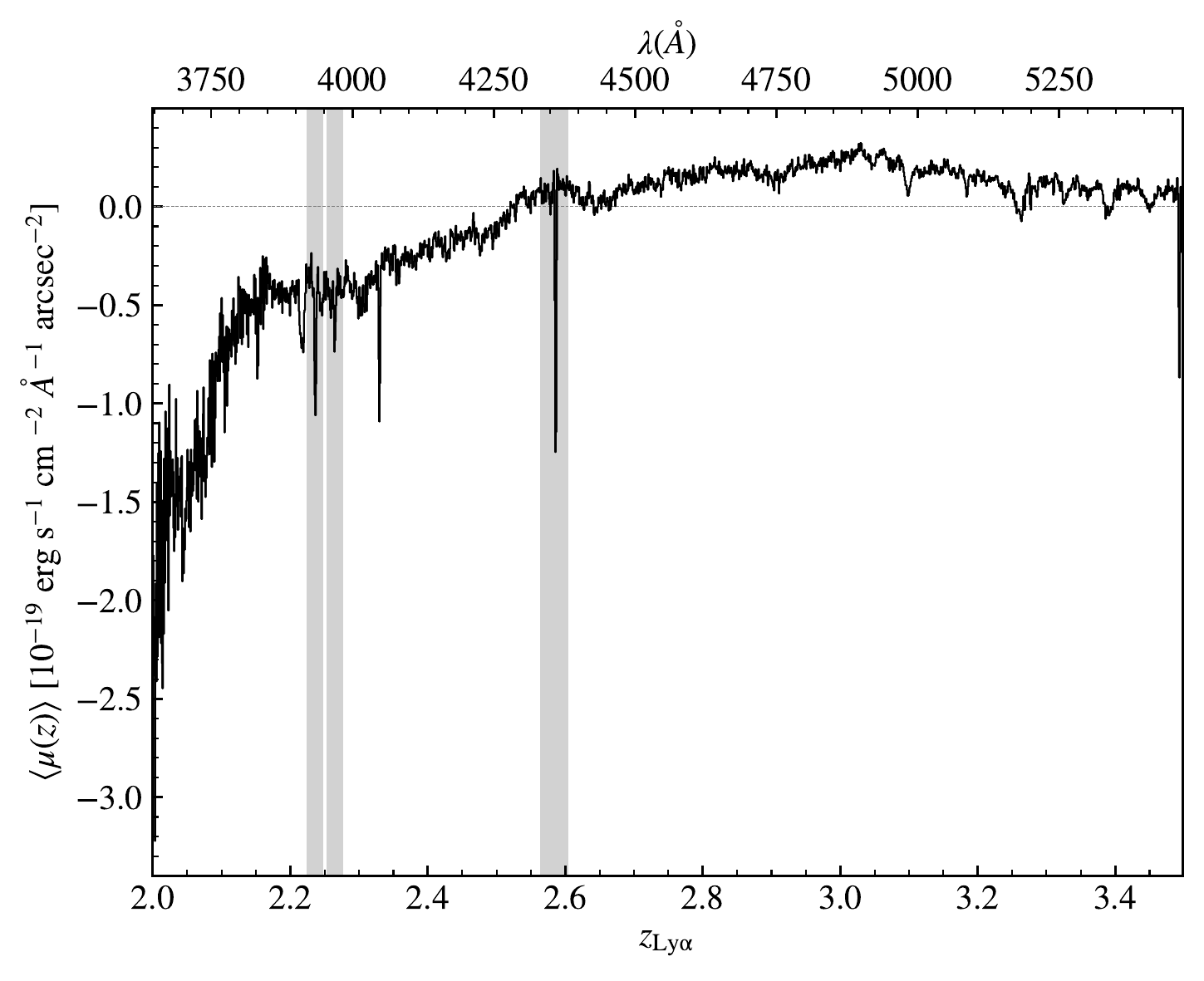}
    \caption{The average residual surface brightness $\langle\mu(z)\rangle$, obtained by averaging all the individual residual spectra in the observed frame after subtraction of the bestfit galaxy model sepctra from the LRG spectra. The gray regions, centered at 3934\AA\ and 3969\AA\ spanning 30\AA, respectively, and 4358\AA\ spanning 40\AA, are masked for zero-redshift Ca H\&K lines and the strong Mercury G line from streetlamps.}
    \label{fig:stack}
\end{figure}

Following \citet{Croft2016}, we estimate the quasar-Ly$\alpha$ emission surface brightness cross-correlation, $\xi_{q\alpha}(r_{\perp},r_\|)$, by summing over all quasar-LRG spectrum pixel pairs separated by $r_\|$ along the line-of-sight direction and $r_{\perp}$ along the transverse direction within a certain bin:
\begin{equation}\label{eq:2dxcf}
\xi_{q \alpha}(r_\|,r_{\perp})=\frac{1}{\sum_{i=1}^{N(\vec{r})} w_{r i}} \sum_{i=1}^{N(\vec{r})} w_{r i} \Delta_{\mu, r i},
\end{equation}
where $N(\vec{r})$ is the number of LRG spectrum pixels within the separation bin centered at the position $\vec{r}=(r_{\perp},r_\|)$ and $\Delta_{\mu,ri}=\mu_{ri}-\langle\mu(z)\rangle$ denotes the fluctuation of Ly$\alpha$ surface brightness for the $i$-th pixel in this bin. Here, $\mu_{ri}$ is the residual surface brightness calculated by subtracting the bestfit galaxy model spectra from the observed LRG spectra and dividing the residuals by the angular area of SDSS fiber, and $\langle\mu(z)\rangle$ is the 
average residual surface brightness at each redshift (Figure~\ref{fig:stack}), obtained by stacking the surface brightness of all residual LRG spectra in the observed frame. \add{The spectral interval $\Delta\log_{10}\lambda=10^{-4}$ (about 69 ${\rm km\, s^{-1}}$ per pixel) in the SDSS spectra is kept when we compute $\langle\mu(z)\rangle$.}
The pixel weight $\omega_{ri}$ is the inverse variance of the flux, $1/\sigma_{ri}^2$, for valid pixels and zero for masked pixels. To avoid the stray light contamination from quasars on the CCD, similar to \citet{Croft2016}, we exclude any LRG spectrum once it is observed within 5 fibers or less away from a quasar fiber (i.e., $\Delta_{\rm fiber}\leq 5$), as discussed in Appendix~\ref{sec:stray light systematics}. A more detailed analysis of the potential contamination in our measurement and the correction to possible systematics are discussed in Appendix~\ref{sec:systematics}. 

Note that the average residual surface brightness shown in Figure~\ref{fig:stack} differs from that in \citet{Croft2016}, mainly due to improved algorithms in flux-calibration and extraction for DR16\footnote{\url{https://www.sdss.org/dr16/spectro/pipeline/\#ChangesforDR16}}. Regardless, the strong features at the zero redshift calcium H and K lines and Mercury G line remain the largest excursions. This difference has little impact on our following analysis, since the residual continuum contributes only to the statistical noise in the measurement, not the signal. 
The feature around 4050\AA\ might be due to one of the sky emission lines, \ion{Hg}{1} 4047\AA. \add{We decide not to mask it,  as we also do not specially deal with regions where other sky lines may reside. Since it is not the flux of $\langle\mu(z)\rangle$ itself but the fluctuation level $\Delta_{\mu}$ relative to it matters (see Equation~\ref{eq:2dxcf}), this feature, shared by all individual residual spectra, will not affect our cross-correlation measurements.}

\begin{figure}
    \centering
    \includegraphics[width=.5\textwidth]{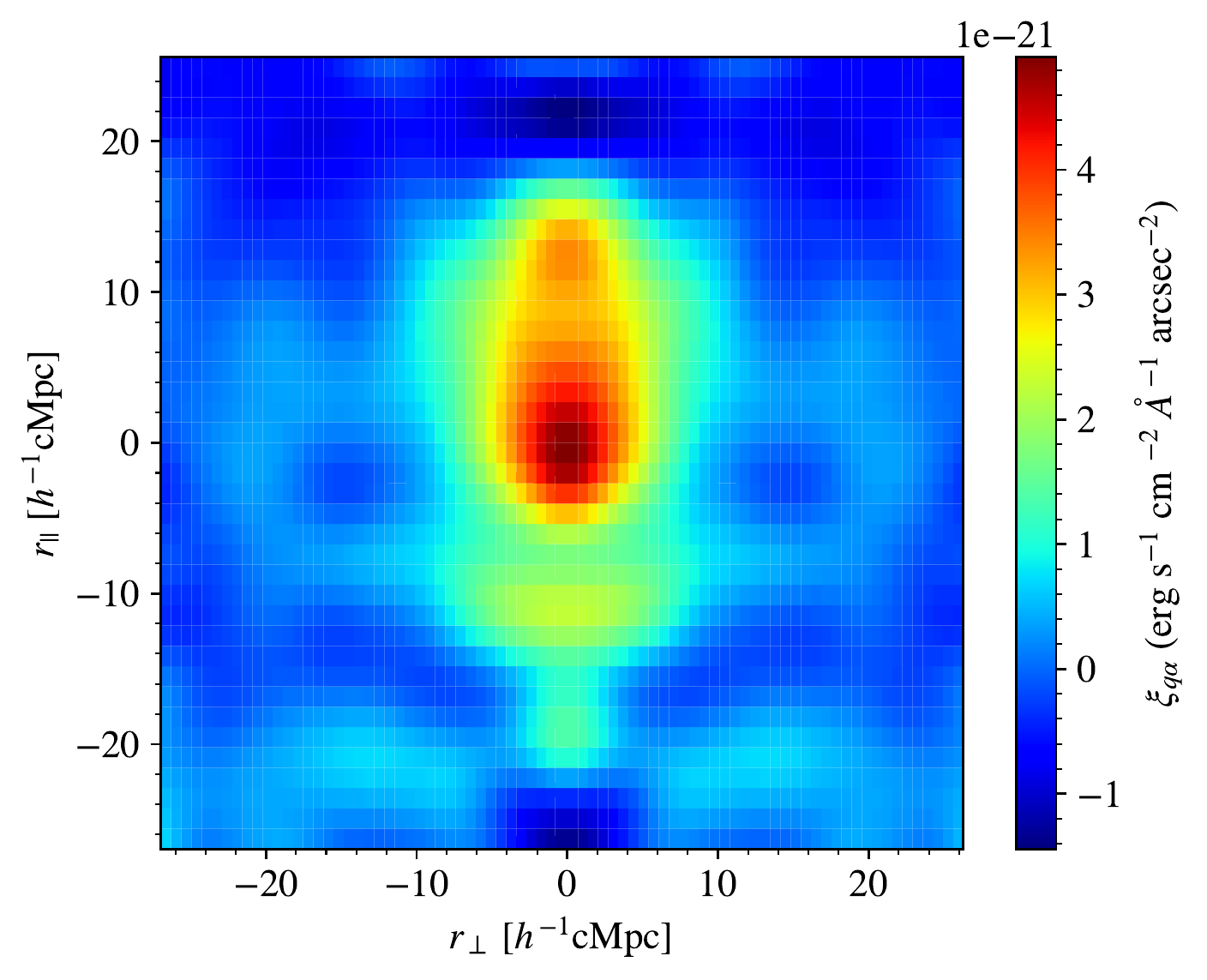}
    \caption{The quasar-Ly$\alpha$ emission cross-correlation as a function of $r_{\perp}$ and $r_\|$. To reduce noise
in the image, the data is smoothed with a 2D Gaussian kernel with a standard deviation of 4 $h^{-1}$cMpc. Potential light contamination is removed by pixel veto. \add{For display, the pattern is mirrored along $r_{\perp}=0$.}}
    \label{fig:linear_xcf_conv}
\end{figure}

We show the quasar-Ly$\alpha$ emission cross-correlation on a linear scale in Figure~\ref{fig:linear_xcf_conv}. The contours are somewhat stretched along the $r_\|$ direction for $r_\perp$ below a few $h^{-1}$Mpc. \cite{Croft2016} quantified the redshift-space anisotropies by assuming a linear $\Lambda$CDM correlation function shape distorted by a peculiar velocity model, which includes standard linear infall for large-scale flows and a small-scale random velocity dispersion. In fact the elongation in the $r_\|$ direction can be caused by a combination of multiple factors, including the intrinsic velocity dispersion of quasars in their host halos, the intrinsic velocity dispersion of the sources of Ly$\alpha$ emission, and quasar redshift uncertainties. 
The uncertainty in quasar redshifts primarily comes from systematics offsets between measured redshifts adopting different indicators, which can sometimes become large due to the complexity of physical processes related to broad emission lines. That makes it difficult to precisely and
accurately distangle a systemic redshift. For example, the variation of quasar redshift offsets between \texttt{Z\_PCA} (redshift estimated by PCA) and \texttt{Z\_Mg\uppercase\expandafter{\romannumeral2}} (redshift indicated by Mg\uppercase\expandafter{\romannumeral2} emission lines) in DR16 can reach over $\pm$ 500 ${\mathrm{km~s^{-1}}}$ \citep{Paris2018,Lyke2020,Brodzeller2022}, which corresponds to $\sim \pm 4.7h^{-1}$cMpc.

\subsection{Projected Ly$\alpha$ Emission Surface Brightness}\label{sec:proj_SB}

In this subsection, we measure the projected Ly$\alpha$ SB profile in a psedo-narrow band  by collapsing the 2D cross-correlation along the line-of-sight direction. 
There are previous studies of \lya\ surface brightness profile around quasars. In order to compare to our derived profile, we first summarize those observations.

\cite{Cai2019} studied quasar circumgalactic Ly$\alpha$ emission using KCWI observations of 16 ultraluminous Type I QSOs at $z = 2.1-2.3$. They integrated over a fixed velocity range of $\pm$1000 km s$^{-1}$ around the centroid of Ly$\alpha$ nebular emission to calculate the SB. The median Ly$\alpha$ SB profile in their work can be described by the following power-law profile centered at the QSO at projected radius $r_\perp$ of 15-70 pkpc, which we denote as ${\rm SB_C}$:
\begin{equation}\label{sb_c}
\begin{aligned}
\mathrm{SB}_{\mathrm{C}}(z \approx 2.3) & = 3.7 \times 10^{-17} \times(r_\perp / 10\ \mathrm{pkpc})^{-1.8}  \\
 & \mathrm{erg}\  \mathrm{s}^{-1} \mathrm{~cm}^{-2} \operatorname{arcsec}^{-2}.
\end{aligned}
\end{equation}

\cite{Borisova2016} found large Ly$\alpha$ nebulae on the spatial extent of $>$100 pkpc from a MUSE snapshot survey on 17 \add{radio quiet} QSOs at $z>3.1$. \add{Twelve of them are selected specifically for their study from the catalog of \cite{Veron2010}, as the brightest radio-quiet quasars known in the redshift range of $z=3.0-3.3$, and the other five at $z=3.6-4.0$ are selected originally for studying absorption line systems in quasar spectra.} They fixed the width of their pseudo-NB images to the maximum spectral width of the Ly$\alpha$ nebulae, with a median of 43\AA. The median of their integrated SB profiles, denoted as ${\rm SB}_{\rm B}$ here, can be described as:
\begin{equation}\label{sb_b}
\begin{aligned}
\mathrm{SB}_{\mathrm{B}}(z \approx 3.1) &= 3.2 \times 10^{-17} \times(r_\perp / 10\ \mathrm{pkpc})^{-1.8} \\
& \operatorname{erg} \mathrm{s}^{-1} \mathrm{~cm}^{-2} \operatorname{arcsec}^{-2}.
\end{aligned}
\end{equation}

Besides, \cite{Croft2018} used a power-law,
\begin{equation}
\begin{aligned}
 {\rm SB_{Croft}}(z\approx2.55) &=  3.5 \times 10^{-19} \times \left( r_\perp/{\rm cMpc}\right)^{-1.5 } \\ 
 & {\rm erg}\ {\rm s}^{-1} {\rm cm}^{-2} {\rm arcsec}^{-2},
\end{aligned}
\end{equation}
\add{ to follow the broad trend seen in the data}.

If we make a simple correction for cosmological surface brightness dimming to $z=2.40$, the median redshift of our quasar sample, by scaling with a factor of $(1 + z)^4$,
the above SB profiles become
\begin{equation}\label{Eq:corrected_sb_bc}
\begin{aligned}
\mathrm{SB}_{\mathrm{C}}(z \approx 2.40) =& 3.3\times 10^{-17} \times(r_\perp / 10\ \mathrm{pkpc})^{-1.8} \\
& \mathrm{erg}\ \mathrm{s}^{-1} \mathrm{~cm}^{-2} \operatorname{arcsec}^{-2}, \\
\mathrm{SB}_{\mathrm{B}}(z \approx 2.40) =& 6.8 \times 10^{-17} \times(r_\perp / 10\ \mathrm{pkpc})^{-1.8}  \\
& \operatorname{erg} \mathrm{s}^{-1} \mathrm{~cm}^{-2} \operatorname{arcsec}^{-2},
\end{aligned}
\end{equation}
and
\begin{equation}\label{Eq:corrected_sb_croft}
\begin{aligned}
 {\rm SB_{Croft}}(z\approx2.40) &=  4.16 \times 10^{-19} \times \left( r_\perp/{\rm cMpc}\right)^{-1.5 }\\
 &{\rm erg}\ {\rm s}^{-1} {\rm cm}^{-2} {\rm arcsec}^{-2}.
\end{aligned}
\end{equation}

To properly compare our measured SB with these previous work, 
we firstly collapse the 2D cross-correlation measurement in Section~\ref{sec:xiqa} along $r_\|$ to obtain the SB as a function of $r_{\perp}$. We integrate the cross-correlation over a fixed line-of-sight window of $\pm 1000$ km s$^{-1}$, corresponding to 
a window spanning $\pm 4$\AA\ around $\lambda_{{\rm Ly}\alpha}\approx1216$\AA\ in the $z=2.40$ quasar rest frame, or a window of $\pm 9.37 h^{-1}{\rm cMpc}$ around the quasar.

We use the jackknife method to compute the standard deviation of the obtained SB, by drawing a jackknife sample set from the 885 LRG subsamples and perform a cross-correlation with the quasar sample. The covariance matrix $C_{ij}$ can be written as:
\begin{equation}
\begin{aligned}
&C_{i j}(r_{\perp,i},r_{\perp,j})=\frac{n-1}{n}\times \\
&\sum_{k=1}^{n}\left[{\rm SB}_k\left(r_{\perp,i}\right)-\overline{\rm SB}\left(r_{\perp, i}\right)\right]\left[{\rm SB}_k\left(r_{\perp,j}\right)-\overline{\rm SB}\left(r_{\perp,j}\right)\right],
\end{aligned}
\end{equation}
where ${\rm SB}_k\left(r_{\perp,i}\right)$ is the surface brightness in bin $i$ centered at the transverse separation $r_{\perp,i}$ for the jackknife sample $k$, $\overline{\rm SB}(r_{\perp,i})$ denotes the surface brightness measured from the full LRG data set, and the number of jackknife samples, $n$, is 885.

\begin{figure}
    \centering
    \includegraphics[width=0.475\textwidth]{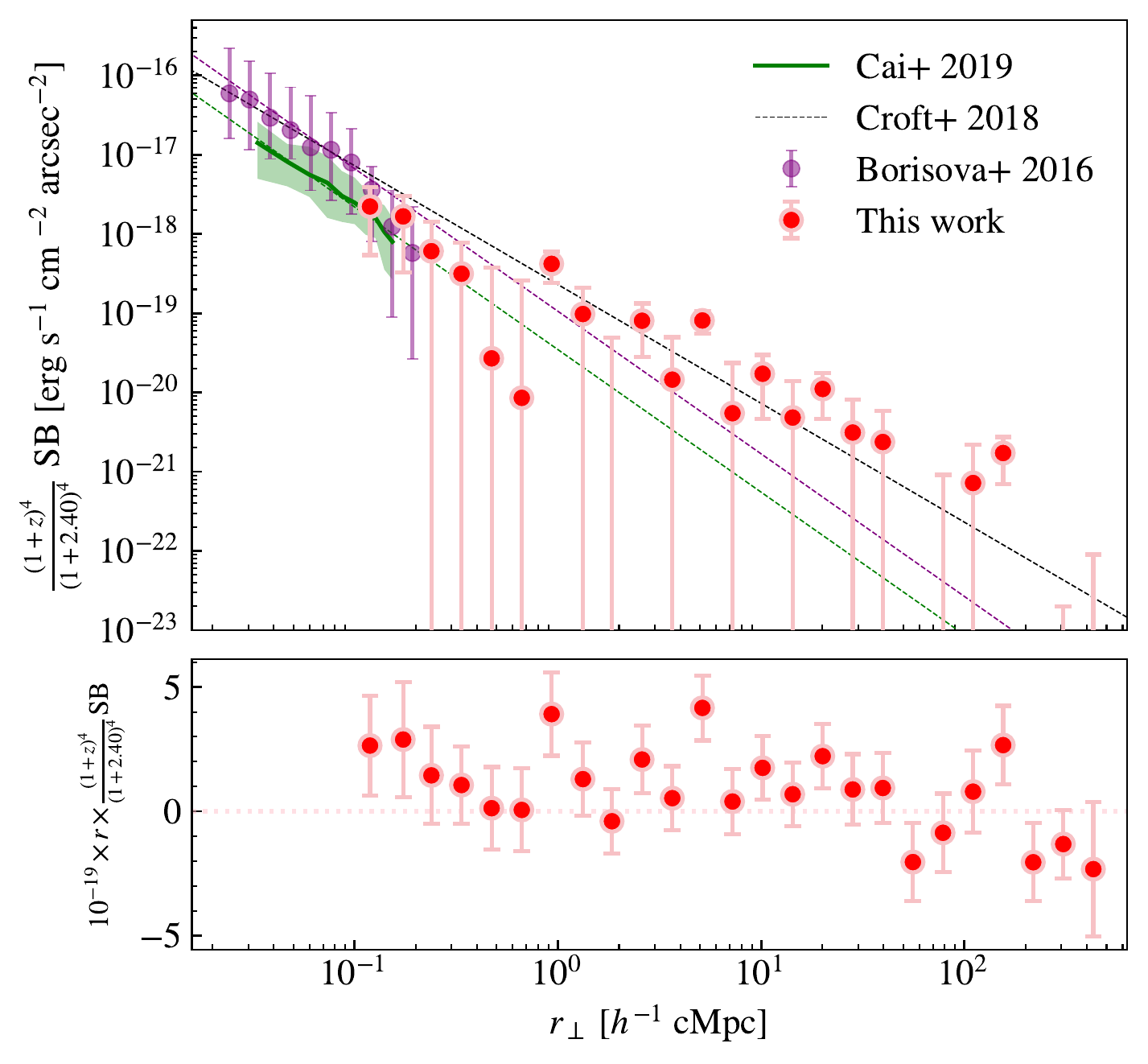}
    \caption{
    Projected \lya\ surface brightness profile (red points) around quasars obtained from our cross-correlation measurement. For comparison, the power-law fit (Equation~\ref{Eq:corrected_sb_croft}) from the intensity mapping result in \citet{Croft2018} is shown as the black dashed line. The SB profiles from observations of \lya\ emission around quasars on smaller scales are shown as the green shaded region (representing the range of 25th and 75th percentiles in \citealt{Cai2019}) and purple points \citep{Borisova2016}, with the green and purple dashed lines denoting the power-law fit and extrapolation (Equation~\ref{Eq:corrected_sb_bc}).
    \add{In the bottom panel, the measured SB is shown in linear scale.}
    }
    \label{fig:SB_profile}
\end{figure}

As shown in Figure~\ref{fig:SB_profile}, we have a detection of the SB profile at projected radius $r_\perp$ ranging from $\sim$0.1$h^{-1}$cMpc to $\sim$100$h^{-1}$cMpc.

The SB profile within $r_{\perp}\leq 0.5\ h^{-1} {\rm cMpc}$ appears to be consistent with the observations of the QSO nebulae on smaller scales in \cite{Cai2019} and \cite{Borisova2016}, 
and on scales of $1 h^{-1}{\rm cMpc} \leq r_{\perp} \leq 10 h^{-1}{\rm cMpc} $ our profile broadly agrees with the power-law fit in \cite{Croft2018}.

\subsection{Multipoles of the Redshift-Space Two-Point Correlation Function}\label{sec:2pcf}

In addition to measure the quasar-\lya\ emission cross-correlation function (a.k.a. two-point correlation function; 2PCF) in bins of $r_\perp$ and $r_\|$, to better describe its shape, we further measure the cross-correlation in bins of $s$ and $\mu$, where $s$ is the separation between quasars and Ly$\alpha$ pixels, i.e.,  $s=\sqrt{r_{\perp}^2+r_\|^2}$, and $\mu$ is cosine of the angle between $\vec{s}$ and
the line-of-sight direction, $\mu=r_\|/s$.

The redshift-space 2PFC $\xi(s,\mu)$ can be expanded into multipoles, with the multipole moment $\xi_\ell$ calculated by \citep{Hamilton1992}:

\begin{equation}\label{eq: xi_l,general}
\xi_{l}(s) =\frac{2 l+1}{2} \int_{-1}^{1} \xi(s, \mu) \mathcal{L}_\ell(\mu) \mathrm{d} \mu, \\
\end{equation}
where $\mathcal{L}_\ell$ is the $\ell$-th order Legendre polynomial. In the linear regime \citep{Kaiser1987}, there are three non-zero components of the redshift-space 2PCF: the monopole $\xi_0$, the quadrupole $\xi_2$ and the hexadecapole $\xi_4$,
\begin{equation} \label{eq:xi_rmu,general}
    \xi(s,\mu) = \sum_{\ell=0,2,4}\xi_\ell(s) \mathcal{L}_\ell(\mu).
\end{equation}

At small transverse separations, however, the redshift-space 2PCF is affected by the small-scale non-linear effect, such as the Finger-of-God (FoG) effect, and also the quasar redshift uncertainty in our cases. To reduce the small-scale contamination, we follow \citet{Kevin2019} to adopt the truncated forms of the multipoles by limiting the calculation to large transverse separations ($r_{\perp}>r_{\perp,{\rm cut}}$), 
\begin{equation}\label{eq:modified multipole}
    \hat{\xi}_\ell = \frac{2\ell+1}{2}\int_{-\mu_{\max}}^{\mu_{\max}}\xi(s,\mu) \mathcal{L}_\ell(\mu)d\mu,
\end{equation}
where $\mu_{\max}=\sqrt{1-(r_{\perp,\mathrm{cut}}/s)^2}$. The transformation between $\mathbf{\xi}=\left(\xi_0,\xi_2,\xi_4\right)^{T}$ and  $\mathbf{\hat{\xi}}=\left(\hat{\xi}_0,\hat{\xi}_2,\hat{\xi}_4\right)^{T}$ can be described using a $3\times3$ matrix $\mathbf{R}$:
\begin{equation}\label{eq:hat xi}
    \mathbf{\hat{\xi}} = \mathbf{R} \mathbf{\xi},
\end{equation}
where 
\begin{equation}\label{eq:R_lk}
    R_{\ell k} = \frac{2\ell+1}{2}\int_{-\mu_{\max}}^{\mu_{\max}}  \mathcal{L}_\ell(\mu)\mathcal{L}_k(\mu)d\mu \quad {\rm for ~} \ell,k=0,2,4.
\end{equation}
In our measurement we set $r_{\perp,{\rm cut}}=4h^{-1}{\rm cMpc}$ to ensure that bulk of small-scale contamination is excluded. The multipole measurements will be presented along with the modeling results.

\subsection{Modeling the Quasar-Ly$\alpha$ Emission Cross-correlation}\label{sec:model}

\add{In \cite{Croft2016}, the amplitude of the measured quasar-\lya\ emission cross-correlation, if modelled by relating \lya\ emission to star-forming galaxies, would imply a value of \lya\ emissivity to be comparable to that inferred from the cosmic SFRD without dust correction,  appearing too high compared with the predictions from the \lya\ luminosity functions (LF) of \lya\ emitting galaxies.}
In \citet{Croft2018}, with the correction to the systematic effect from quasar clustering and the complementary measurement of Ly$\alpha$ forest-Ly$\alpha$ emission cross-correlation, the detected \lya\ emission is found to be explained by \lya\ emission associated with quasars based on populating a large hydrodynamic cosmological simulation. 
In this subsection we will revisit both scenarios 
by constructing a simple analytic model to describe the measured Ly$\alpha$ intensity, and argue that the observed Ly$\alpha$ emission cannot be only contributed by quasars. The simple model can also be applied to  Ly$\alpha$ forest-Ly$\alpha$ emission cross-correlation, and our corresponding prediction and detailed analysis are presented in Section~\ref{sec:forest correlation}.

We assume that the Ly$\alpha$ emission from sources clustered with quasars contribute the bulk of the detected signals on large scales, while on small scales the Ly$\alpha$ photons associate with the central quasar count. Supposing that $\langle\mu_\alpha\rangle$ is the mean surface brightness of Ly$\alpha$ emission, $b_q$ and $b_{\alpha}$ are the linear bias factors of quasars and Ly$\alpha$ sources, respectively, in the linear regime the non-vanishing multipoles of the redshift-space quasar-\lya\ emission cross-correlation are given by
\begin{equation}\label{eq:model galaxy multipoles}
\begin{aligned}
&{\xi}_{0}(s)= b_q b_{\alpha} \langle\mu_\alpha\rangle f_{\beta,0} {\xi}_{mm}(r), \\
&{\xi}_{2}(s)=b_q b_{\alpha} \langle\mu_\alpha\rangle f_{\beta,2} \left[{\xi}_{mm}(r)-\bar{\xi}_{mm}(r)\right], \\
&{\xi}_{4}(s)=b_q b_{\alpha} \langle\mu_\alpha\rangle f_{\beta,4} \left[\xi_{mm}(r)+\frac{5}{2} \bar{\xi}_{mm}(r)-\frac{7}{2} \bar{\bar{\xi}}_{mm}(r)\right],
\end{aligned}
\end{equation}
where \citep[e.g.,][]{Percival2009}
\begin{equation}\label{eq:model galaxy f}
\begin{aligned}
f_{\beta,0} &=  1+\frac{1}{3}\left(\beta_{q}+{\beta}_{\alpha}\right)+\frac{1}{5} \beta_{q} {\beta}_{\alpha},  \\
f_{\beta,2} &=  \frac{2}{3}\left(\beta_{q}+{\beta}_{\alpha}\right)+\frac{4}{7} \beta_{q} {\beta}_{\alpha}, \\
f_{\beta,4} &= \frac{8}{35} \beta_{q} {\beta}_{\alpha},
\end{aligned}    
\end{equation}
and \citep[e.g.,][]{Hawkins2003}
\begin{equation}
\begin{aligned}
&\bar{\xi}(r)=\frac{3}{r^{3}} \int_{0}^{r} \xi\left(r^{\prime}\right) r^{\prime 2} \mathrm{~d} r^{\prime}, \\
&\bar{\bar{\xi}}(r)=\frac{5}{r^{5}} \int_{0}^{r} \xi\left(r^{\prime}\right) r^{\prime 4} \mathrm{~d} r^{\prime}.
\end{aligned}
\end{equation}
Note that $r$ is for the distance in the real space and $s$ denotes the distance in the redshift space and in the above expressions $r=s$. Then the model for the truncated two-point correlation function $\hat{\xi}$ can be obtained according to Equations~(\ref{eq:hat xi}) and (\ref{eq:R_lk}).

The redshift-space distortion parameter $\beta_q$ for quasars depicts the redshift-space anisotropy caused by peculiar velocity, $\beta_q=\Omega_{m}^{0.55}(z=2.4)/b_q$. We fix $b_q=3.64$ according to \cite{Font-Ribera2013}. The redshift-space distortion parameter $\beta_\alpha$ for \lya\ emission is similarly defined. We set $b_\alpha = b_q$ for the case that the main contributors to \lya\ emission are clustered quasars and $b_\alpha=3$ for the case that \lya\ emission is dominated by contributions from star-forming galaxies. 
\add{
A value of 3 appears to be a good estimate of the luminosity-weighted bias $b_\alpha$ for star-forming galaxies. Following \citet{Croft2016}, we find that $b_\alpha$ is within $\sim$5\% of 3 with different low halo mass cuts and different prescriptions of the stellar mass-halo mass relation at $z\sim 2.4$ \citep[e.g.,][]{Moster2010,Moster2013,Behroozi2019}.
}
In both scenarios we leave $\beta_\alpha$ and $\langle\mu_\alpha\rangle$ as free parameters to be fitted. We note that $\beta_\alpha$ can potentially include additional effects other than the Kaiser effect, such as the Ly$\alpha$ radiative transfer on clustering \citep{Zheng2011a}. 

We also model the \lya\ SB profile. As discussed in Section~\ref{sec:proj_SB}, previous observations indicate that the small-scale SB profile can be well described by a power law with an index of $-1.8$. We therefore decompose the full SB profile into two components: the one-halo term ${\rm SB_{1h}}$ dominated by \lya\ emission associated with the central quasars and the two halo-term ${\rm SB_{2h}}$ by the clustered Ly$\alpha$ sources,
\begin{equation}\label{eq:model galaxy SB}
\begin{aligned}
    {\rm SB_{1h}} &= {\rm SB_0} \left(\frac{r_{\perp}}{1 h^{-1}{\rm cMpc}}\right)^{-1.8},\\
\mathrm{SB}_{2 \mathrm{h}} 
&=\frac{\rho_{\rm Ly\alpha}}{4 \pi(1+z)^{2}} \int_{\pi_{\rm min}}^{\pi_{\rm max}} \xi(r_\perp,r_{\|}) dr_{\|}\\ 
&= \frac{\rho_{\rm Ly\alpha}}{4 \pi(1+z)^{2}} b_q b_\alpha \left(f_{\beta, 0} w_{p, 0}+f_{\beta, 2} w_{p, 2}+f_{\beta, 4} w_{p, 4}\right).
\end{aligned}
\end{equation}
Here $\xi$ is the linear correlation function between quasars and \lya\ emission sources (quasars or star-forming galaxies) in redshift space, 
$\rho_{\rm Ly\alpha} =  4\pi \langle\mu_\alpha\rangle [H(z)/c] \lambda_\alpha (1+z)^2 $ is the comoving Ly$\alpha$ luminosity density \citep{Croft2016}, $\pi_{\max}$ and $\pi_{\min}$ correspond to $\pm$9.37$h^{-1}$cMpc, the width of the pseudo-narrow band used in \S~\ref{sec:proj_SB}. The projected cross-correlation function is put in the form of the projected multipoles, which are calculated as

\begin{equation}
\begin{aligned}
w_{p, 0}(r_\perp) &=\int_{\pi_{\min }}^{\pi_{\max }} \xi_{mm}(r) \mathcal{L}_{0}(\mu) d r_{\|}, \\
w_{p, 2}(r_\perp)  &=\int_{\pi_{\min }}^{\pi_{\max }}\left[{\xi}_{mm}(r)-\bar{\xi}_{mm}(r)\right] \mathcal{L}_{2}(\mu) d r_{\|}, \\
w_{p, 4}(r_\perp)  &=\int_{\pi_{\min }}^{\pi_{\max }}\left[\xi_{mm}(r)+\frac{5}{2} \bar{\xi}_{mm}(r)-\frac{7}{2} \bar{\bar{\xi}}_{mm}(r)\right]\mathcal{L}_{4}(\mu) d r_{\|}.
\end{aligned}
\end{equation}
with $r=\sqrt{r_\perp^2+r_\|^2}$ and $\mu=r_\|/r$.

With three free parameters (${\rm SB_0}$, $\beta_\alpha$, and $\langle\mu_\alpha\rangle$), we perform a joint fit to the three (large-scale) multipoles and the projected SB profile, assuming that the Ly$\alpha$ sources in the model are mainly quasars and star-forming galaxies, respectively, as discussed in Section~\ref{sec:model galaxy} and Section~\ref{sec:model quasar}.

\subsubsection{Star-forming Galaxies as Ly$\alpha$ Sources}\label{sec:model galaxy}

\begin{figure*}
    \centering
    \includegraphics[width=\textwidth]{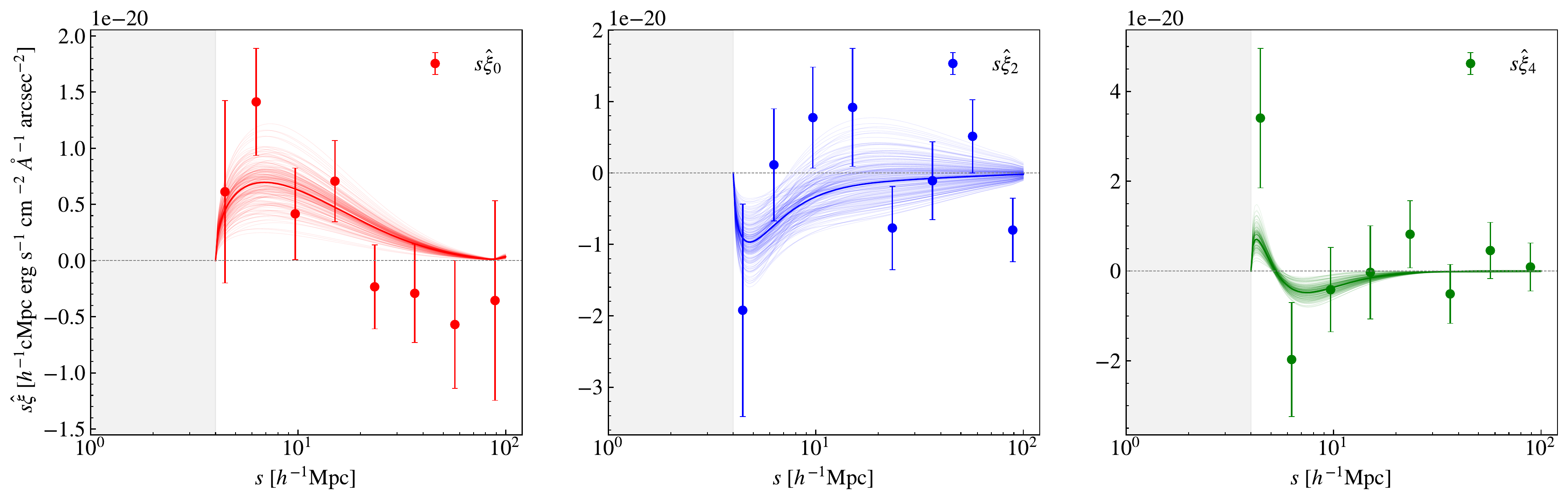}
    \caption{Modified monopole, quadrupole and hexadecapole of the quasar-Ly$\alpha$ emission cross-correlation (see Equation~\ref{eq:modified multipole}) and their fitting results based on the galaxy-dominated model (see Section~\ref{sec:model galaxy}). The points represent our measurements with jackknife error bars. The solid curves denote modelled modified multipoles with parameters randomly drawn from their posterior probability distributions, among which the thickest ones correspond to the best-fits. The modified multipoles remove any information within $r<r_{\perp,{\rm cut}}$, i.e., the gray-shaded regions, to avoid small-scale contamination.
    }
    \label{fig:xi_qa}
\end{figure*}

\begin{figure}
    \centering
    \includegraphics[width=.5\textwidth]{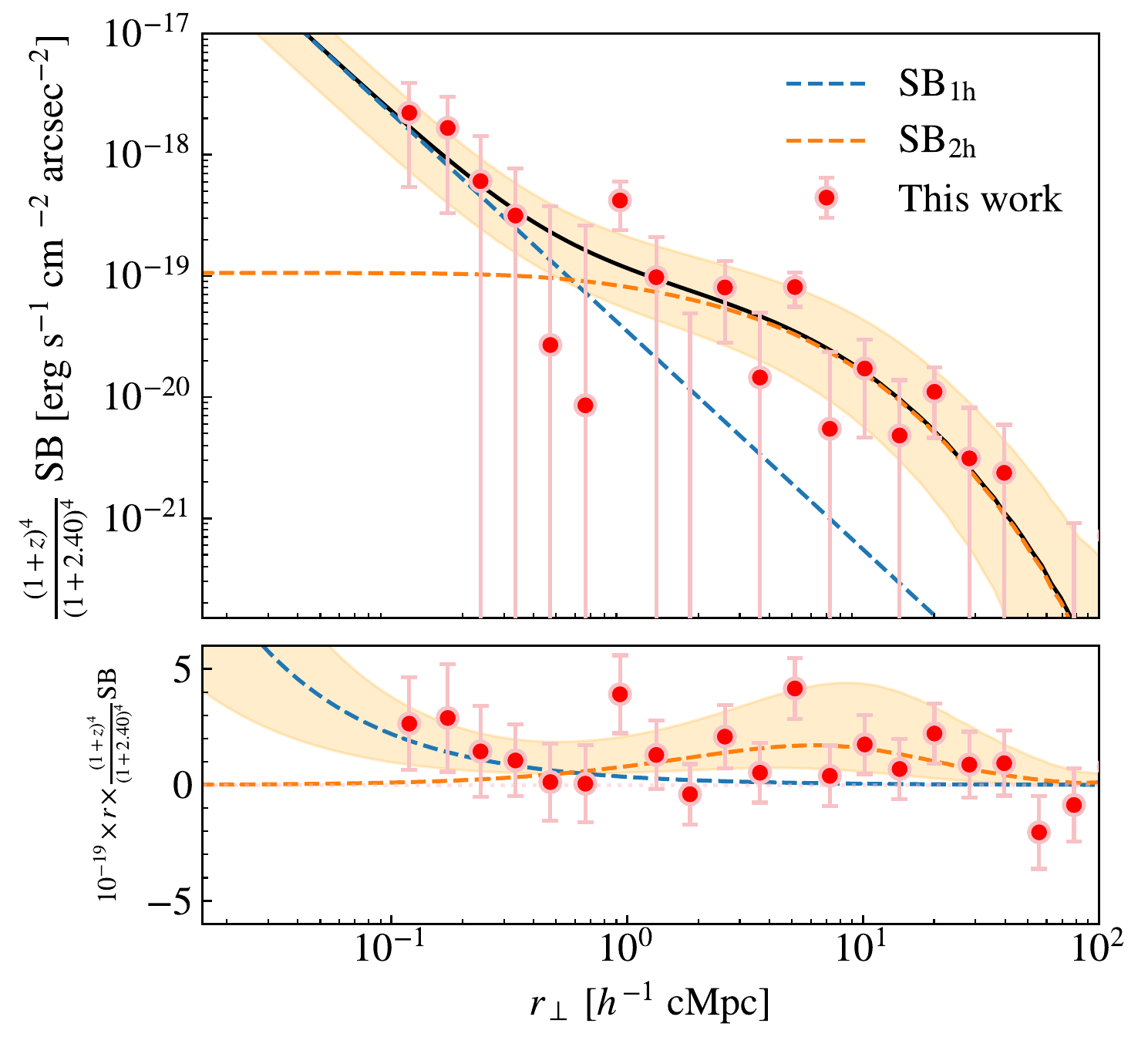}
    \caption{\lya\ SB profile. The data points are from integrating the measured quasar-\lya\ emission cross-correlation function along the line of sight, and the solid curve is the best-fit SB profile for the galaxy-dominated model depicted in Section~\ref{sec:model galaxy}. The dashed lines denote the best-fit one-halo and two-halo term, respectively, and the shaded region represents the $\pm 1\sigma$ range.
    \add{In the bottom panel, linear scale is used in the $y$-axis.}
    }
    \label{fig:MCMC fit SB}
\end{figure}

\begin{figure*}
    \centering
    \includegraphics[width=\textwidth]{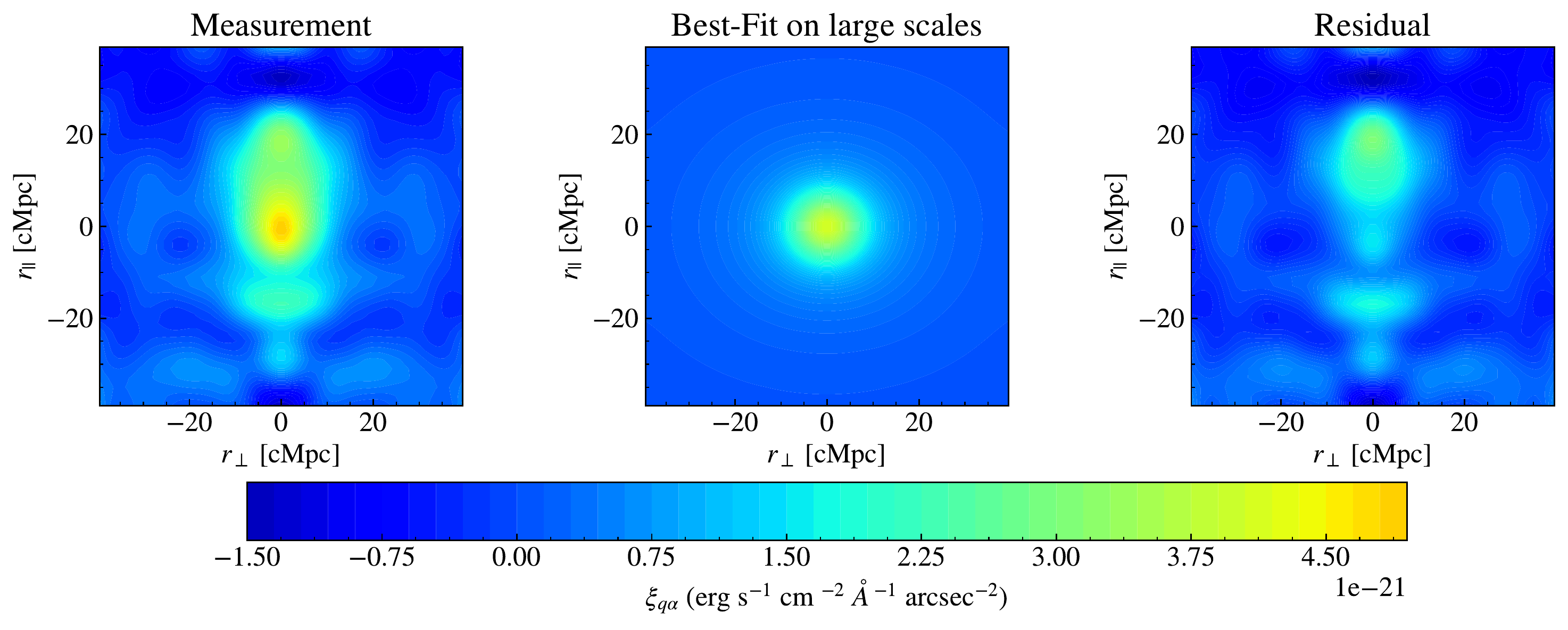}
    \caption{The measured (left panel), best-fit (middle panel) and residual (right panel) quasar-Ly$\alpha$ emission cross-correlation as a function of $r_\|$ and $r_{\perp}$. The model fit is only to large-scale signals by using the modified multipoles (Equation~\ref{eq:modified multipole}). The best-fit pattern shown here is reconstructed from the corresponding multipoles (Equation~\ref{eq:model galaxy multipoles}) with the best-fit parameters. The residual is obtained by subtracting the best-fit model from the measurement, with elongated distortion along the $r_\|$ direction on small scales, the small-scale anisotropy not included in our model. All the three images are smoothed using a 2D Gaussian kernel with a standard deviation of 4$h^{-1}$cMpc. 
    }
    \label{fig:reconstruct 2D}
\end{figure*}

In the case that \lya\ emission is dominated by the contribution from galaxies, we fix $b_\alpha=3$. The best-fit results for the multipoles and the SB profile are shown in Figure~\ref{fig:xi_qa} and Figure~\ref{fig:MCMC fit SB}. Given the uncertainties in the measurements, the model provides a reasonable fit and shows a broad agreement with the trend in the data. The middle panel of Figure~\ref{fig:reconstruct 2D} shows a reconstructed 2D image of the redshift-space linear cross-correlation function from the best-fit model. If it is subtracted from the measurement (left panel), the residual (right panel) is dominated by the small-scale clustering that we do not model.

\begin{figure}
    \centering
    \includegraphics[width=.5\textwidth]{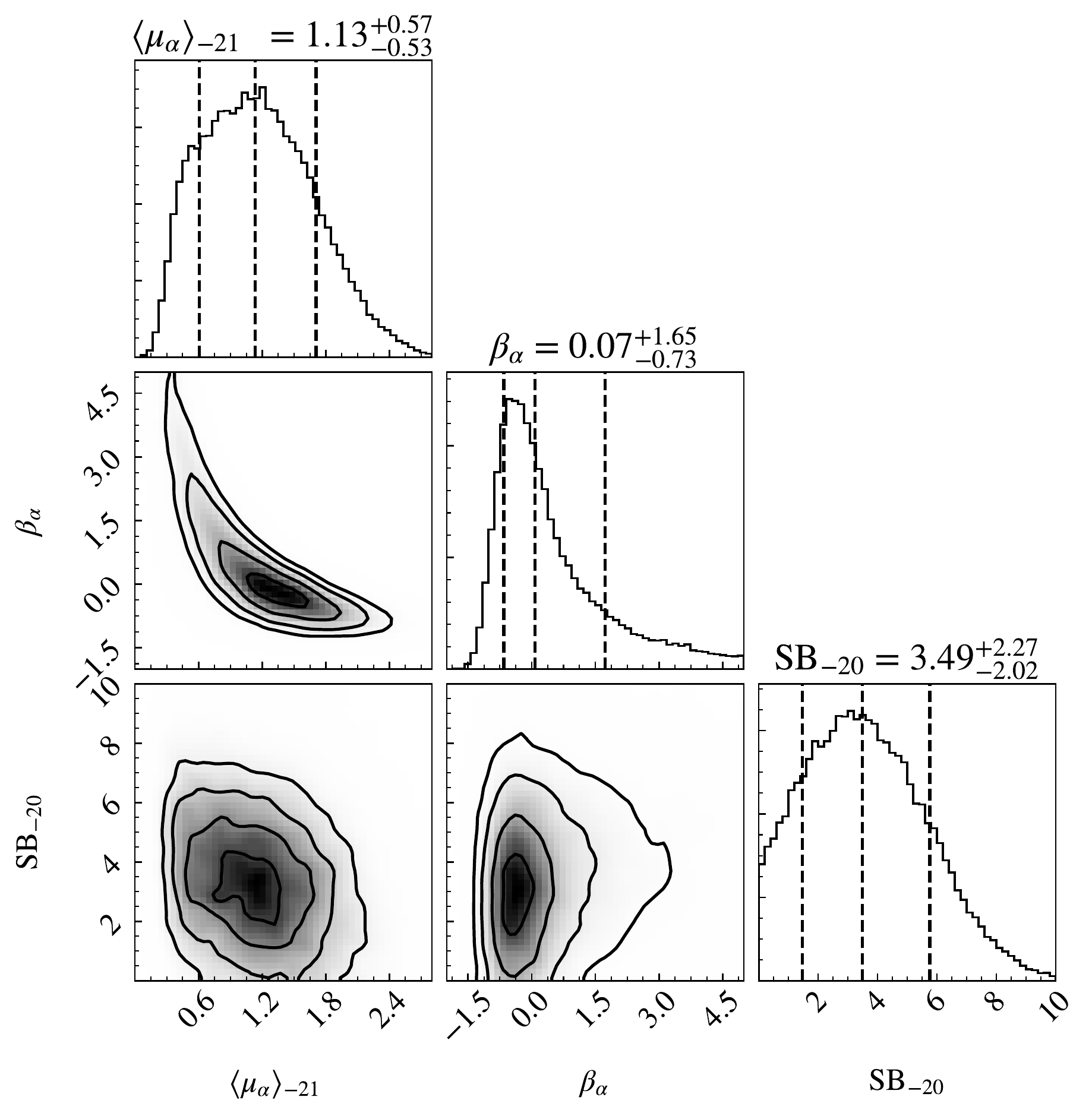}
    \caption{The probability distribution of parameters $\langle\mu_\alpha\rangle$, $\beta_\alpha$ and ${\rm SB_0}$ as a result of the joint fit to the modified multipoles and SB profile of quasar–Ly$\alpha$ emission cross-correlation, with an assumption that star-forming galaxies dominate the large-scale Ly$\alpha$ emission and thus $b_\alpha=3$. $\langle\mu_\alpha\rangle_{-21}$ is $\langle\mu_\alpha\rangle$ in units of $10^{-21}\SBunitsA$, and ${\rm SB}_{-20}$ is ${\rm SB_0}$ in units of $10^{-20}\SBunits$. The dashed lines in the histograms denote 16th, 50th and 84th percentiles of the marginalized distributions.}
    \label{fig:MCMC galaxy model}
\end{figure}

The constraints on the three parameters are presented in Figure~\ref{fig:MCMC galaxy model}. The parameter representing the amplitude of the one-halo term is loosely contrained, ${\rm SB_0}=\valSBzero\times 10^{-20} \SBunits$. The parameter $\langle \mu_\alpha\rangle$, proportional to the comoving \lya\ emissivity or luminosity density, is constrained at the 2$\sigma$ level, $\langle\mu_\alpha\rangle=\valmulya\times 10^{-21} {\rm erg\ s^{-1} cm^{-2} \textup{\AA}^{-1} arcsec^{-2}}$. The redshift-space distortion parameter has a high probability density of being negative but with a tail toward positive values, $\beta_\alpha=\valbetalya$. Given its uncertainty, the value is consistent with that from the Kaiser effect, $\Omega_m(z=2.4)^{0.55}/b_\alpha\simeq 0.32$, and we are not able to tell whether there is any other effect (e.g., caused by radiative transfer; \citealt{Zheng2011a}). 

We note that fitting the clustering measurements leads to an anti-correlation between $\langle \mu_\alpha\rangle$ and $\beta_\alpha$ (Eq.~\ref{eq:model galaxy multipoles} and Eq.~\ref{eq:model galaxy f}; Fig.~\ref{fig:MCMC galaxy model}). If $\beta_\alpha$ is restricted to the formal value of $\sim$0.32 from the Kaiser effect, the constraints on $\langle\mu_\alpha\rangle$ become $1.09_{-0.24}^{+0.25} \times 10^{-21} {\rm erg\ s^{-1} cm^{-2} \textup{\AA}^{-1} arcsec^{-2}}$, a nearly 4$\sigma$ detection. If we set the upper limit of $\beta_\alpha$ to be 0.32 to allow room for radiative transfer effect \citep[e.g.,][]{Zheng2011a}, the constraints change to $\langle\mu_\alpha\rangle=1.44_{-0.38}^{+0.45}\times 10^{-21} {\rm erg\ s^{-1} cm^{-2} \textup{\AA}^{-1} arcsec^{-2}}$. In the following discussions, to be conservative, we take the $\langle \mu_\alpha \rangle$ constraints without these restrictions.

\begin{figure}
    \centering
    \includegraphics[width=.5\textwidth]{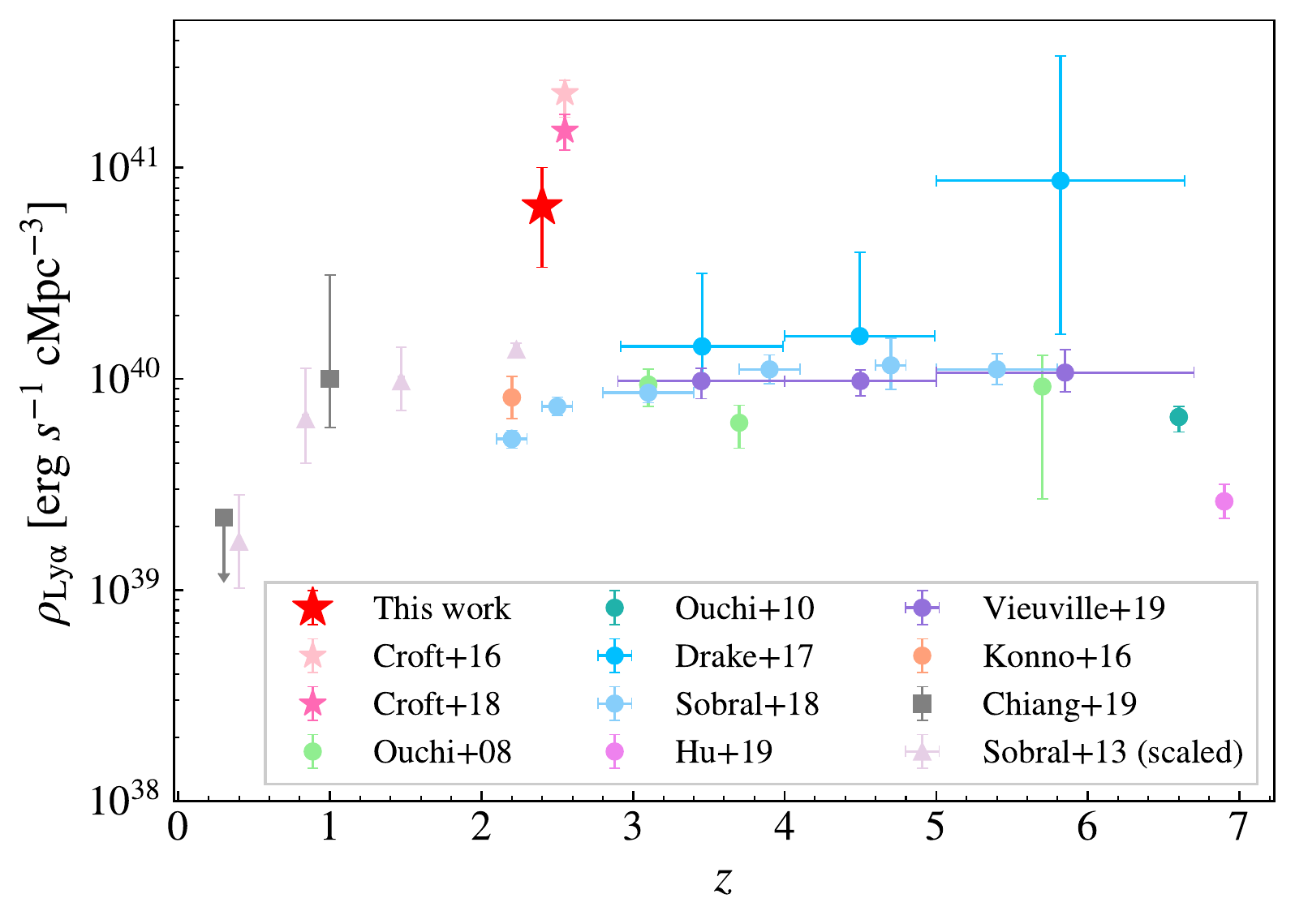}
    \caption{Ly$\alpha$ luminosity density $\rho_{\rm Ly\alpha}$. The red star shows the value inferred from our quasar-\lya\ emission measurement, assuming that the detected Ly$\alpha$ emission is due to star-forming galaxies with a typical luminosity-weighted bias of $b_\alpha=3$. As a comparison, we also show the values with previous intensity mapping measurements \citep{Croft2016,Croft2018,Chiang2019} and those from integrating the \lya\ LFs of LAEs \citep{Ouchi2008,Ouchi2010,Drake2017,Sobral2018,Hu2019,Vieuville2019} \add{and scaling H$\alpha$ luminosities with an escape fraction of 5\% \citep{Sobral2013,Wold2017}}. 
    }
    \label{fig:luminosity density compare}
\end{figure}

The constrained $\langle\mu_\alpha\rangle$ corresponds to a comoving \lya\ luminosity density of $\rho_{\rm Ly\alpha}=\valrholya\times 10^{40} {\rm erg\, s^{-1} cMpc^{-3}}$. This value is about 3.6 times lower than that in \citet{Croft2016} or $\sim 2.2$ times lower than that in \citet{Croft2018}. With the lower amplitude, the fractional uncertainty is larger. The comparison is shown in Figure~\ref{fig:luminosity density compare}. We also show the \lya\ luminosity densities at different redshifts calculated by integrating the \lya\ LFs of LAEs down to low luminosity. For example, LFs in \citet{Ouchi2008,Ouchi2010} are \add{integrated down to 
$L_{\rm Ly\alpha}=0$ with the best-fit Schechter parameters for $z=3.1,3.7,5,7$ and $6.6$;}
that in \citet{Drake2017} down to $\log [L_{\rm Ly\alpha}/({\rm erg\, s^{-1}})] =41.0$; that in \citet{Sobral2018} down to $1.75 \times 10^{41} {\rm erg\, s^{-1}}$. 
\add{These quoted \lya\ luminosity densities are inferred without separating the contribution of potential AGNs except at the very luminous end (see \citealt{Wold2017} for a two-component fit). The luminous end is usually excluded in the parametrized fits to the \lya\ LFs, but they do not contribute much to the total \lya\ luminosity density due to their rather low number density. The quoted LAE \lya\ luminosity densities in Figure~\ref{fig:luminosity density compare} should have included the potential contribution of relatively faint AGNs (with AGNs detected in X-ray and radio contributing at a level of a few percent; \citealt{Sobral2018}). 
}
At $z\sim 2.4$, our inferred \lya\ luminosity density is about one order of magnitude higher than that inferred from the LAE LF, although they can be consistent within the uncertainty. 

\add{We further show the H$\alpha$-converted Ly$\alpha$  luminosity density as done in  \cite{Wold2017}, which is obtained by scaling the H$\alpha$ luminosity density measured in the HiZELS survey \citep{Sobral2013} with an escape fraction of 5\% and a correction about 10\%(15\%) for AGN contribution at $z<1$ ($z>1$). The cosmic \lya\ luminosity density measured by \cite{Chiang2019} through broad-band intensity mapping is also shown, which probes the total background including low surface brightness emission by spatially cross-correlating photons in far-UV and near-UV bands with spectroscopic objects. They claimed that their derived cosmic \lya\ luminosity density \addb{is consistent with} cosmic star formation with an effective escape fraction of 10\%  \addb{assuming that all of the
\lya\ photons originate from star formation}. Combined our measurement with the results of \cite{Chiang2019}, it appears that the cosmic \lya\ luminosity density grows with redshift over $0\lesssim z \lesssim 2.5$, and more data points at different redshifts are expected to confirm this trend.}

\begin{figure}
    \centering
    \includegraphics[width=.5\textwidth]{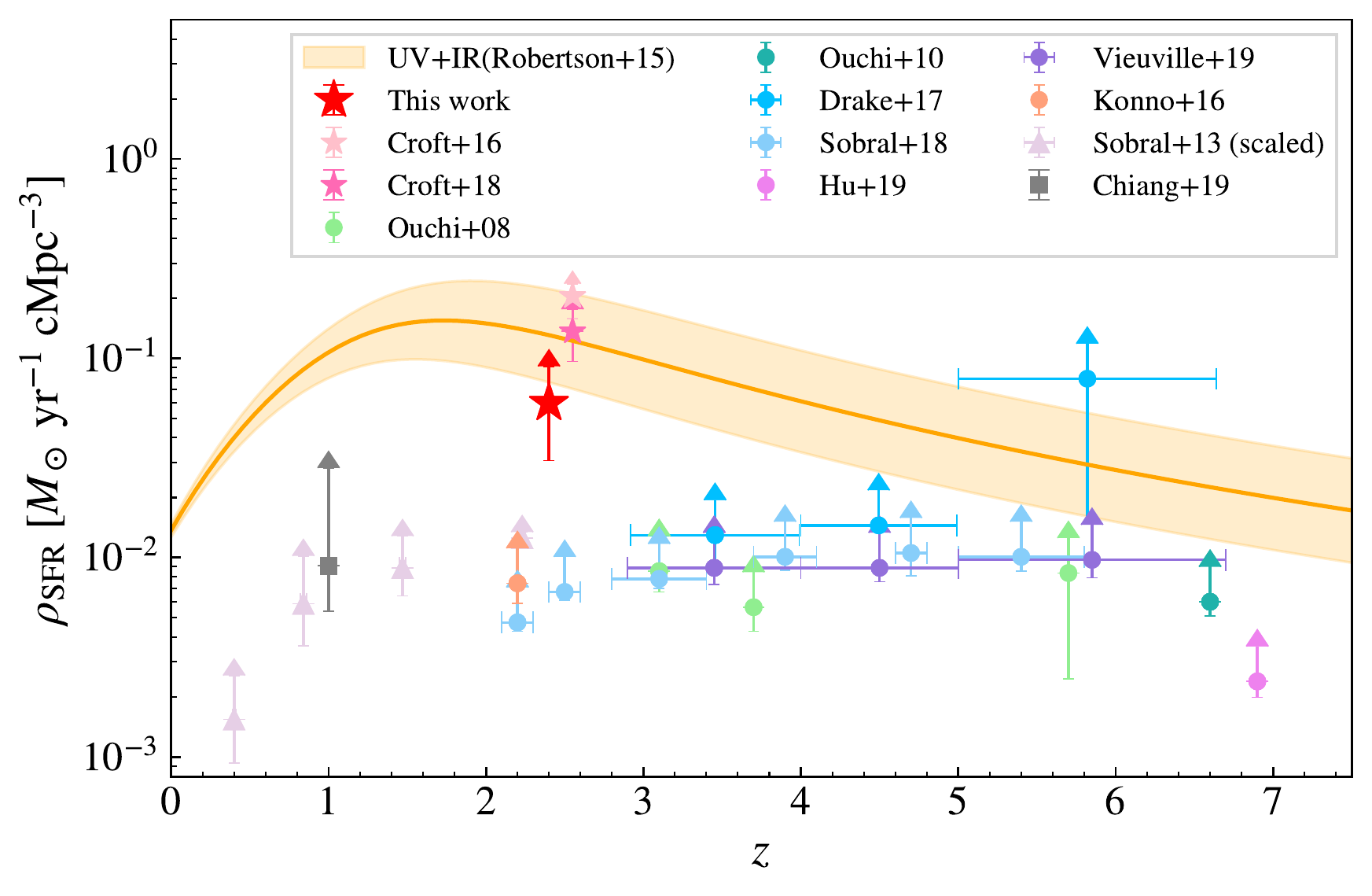}
    \caption{
    Same as Fig.~\ref{fig:luminosity density compare} but the \lya\ luminosity density $\rho_{\rm Ly\alpha}$ is converted to star formation rate density $\rho_{\rm SFR}$ under the assumption that \lya\ emission is purely caused by star formation. \add{Given the effect of dust extinction and \lya\ escape fraction, these \lya-converted $\rho_{\rm SFR}$ values should be considered as lower limits of the intrinsic star formation.}
    The orange shaded region represents the parameterized model for the evolving star formation rate density in \citet{Robertson2015}, based on infrared and ultraviolet observations. }
    \label{fig:SFRD compare}
\end{figure}

If we assume that all the \lya\ emission originates from star formation, we can convert our inferred \lya\ luminosity density to a SFRD, by using a simple conversion \citep{Kennicutt1998},
\begin{equation}
\rho_{\rm SFR}/(M_\odot \mathrm{yr}^{-1} \mathrm{cMpc}^{-3}) = \frac{\rho_{\rm Ly\alpha}/ ({\rm erg ~s^{-1} ~cMpc^{-3}})}{1.1\times 10^{42}({\rm erg\, s^{-1}})/(M_\odot {\rm yr}^{-1})}.
\end{equation}
This gives $\rho_{\rm SFR} = \valSFRD M_\odot \mathrm{yr}^{-1} \mathrm{cMpc}^{-3}$, higher than that from integrating LAE LFs, as shown in Figure~\ref{fig:SFRD compare}. The value is on the low end of the cosmic star formation rate density based on UV and infrared observations \citep[e.g.,][]{Robertson2015}.
\add{However, we emphasize that the \lya-converted $\rho_{\rm SFR}$ in this case should be treated as a lower limit for estimates of the intrinsic star formation, since no correction is applied to account for dust extinction and \lya\ escape fraction. The comparison in Figure~\ref{fig:SFRD compare} is simply to highlight the high amplitude of \lya\ emission inferred from the quasar-\lya\ emission cross-correlation.}

\subsubsection{Quasars as Ly$\alpha$ Sources}\label{sec:model quasar}

In the case that \lya\ emission is dominated by the contribution from quasars, we make a simple assumption that the quasars involved are almost the same, with a typical Ly$\alpha$ luminosity $L_{q,\alpha}$ and a comoving number density $n_q$, so that $\rho_{\rm Ly\alpha} = L_{\alpha,q}n_q$.

We calculate $n_q$ by integrating the luminosity evolution and density evolution (LEDE) model \citep{Ross2013} of the optical quasar luminosity function (QLF), fitted using data from SDSS-\uppercase\expandafter{\romannumeral3} DR9 and allowing luminosity and density to evolve independently. The QLF gives the number density of quasars per unit magnitude, and its integration over the magnitude range from $M_i[z=2]=-30$ to $M_i[z=2]=-18$ yields $n_q\approx\valnq h^3{\rm Mpc}^{-3}$

\begin{figure}
    \centering
    \includegraphics[width=.5\textwidth]{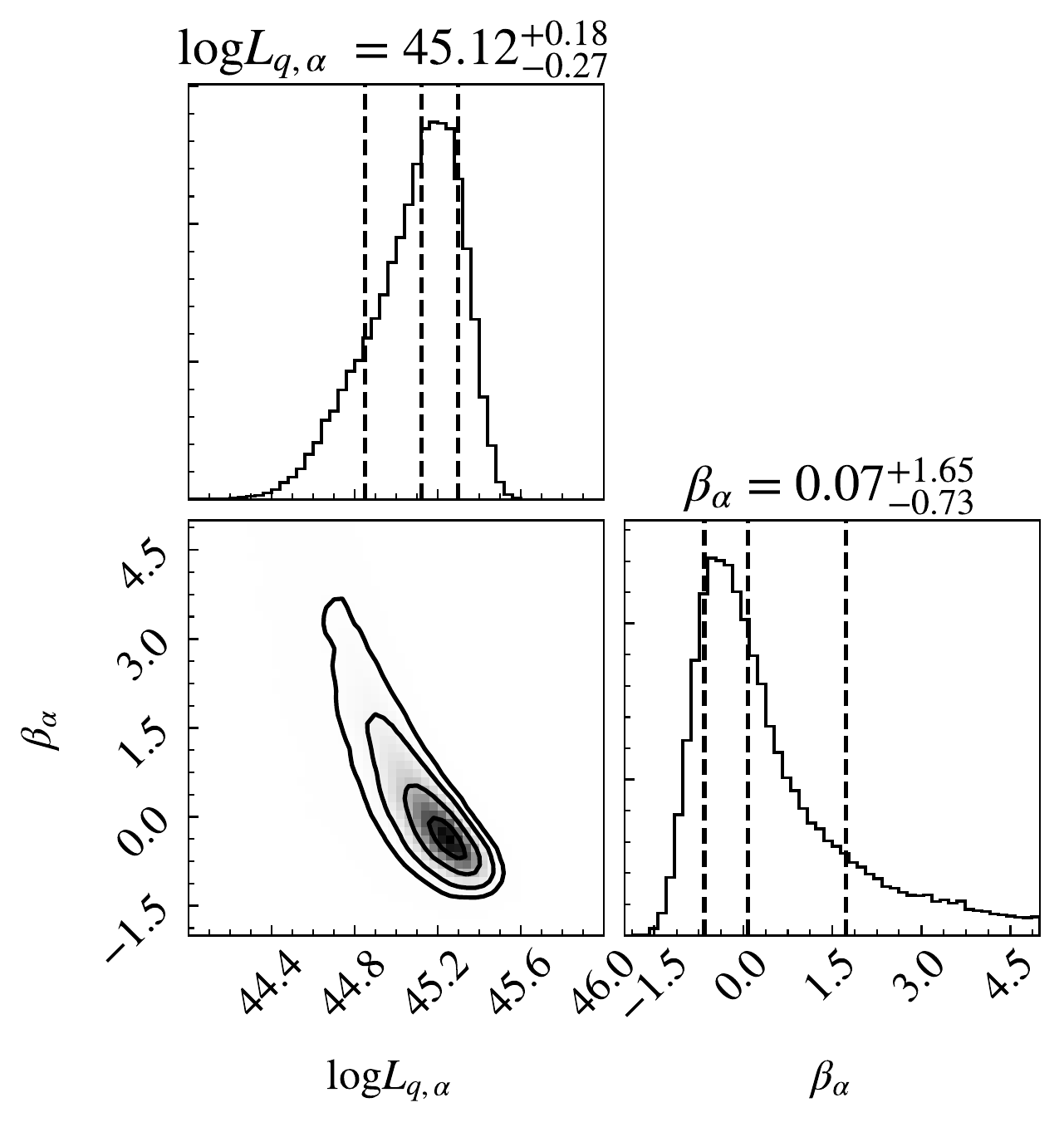}
    \caption{The probability distribution of $L_{q,\alpha}$ and $\beta_{\alpha}$ from the joint fit to the multipoles and SB profiles from the measured quasar-Ly$\alpha$ emission cross-correlation. The mean quasar \lya\ luminosity $L_{q,\alpha}$ is in units of ${\rm erg\, s^{-1}}$. Note that there are actually three parameters, $L_{q,\alpha}$, $\beta_{\alpha}$ and ${\rm SB_0}$, in the model, but here we focus on the constraints on $L_{q,\alpha}$ and $\beta_{\alpha}$. See the text for detail.}
    \label{fig:MCMC quasar model}
\end{figure}

\add{With the analytical model in this quasar-dominant scenario, we jointly fit both the measured cross-correlation multipoles and the SB profile, where $b_\alpha$ is fixed to be $b_q$ and $\rho_{\rm Ly\alpha}$ is interpreted to be $L_{\alpha,q}n_q$, leaving $L_{q,\alpha}$, $\beta_\alpha$ and ${\rm SB_0}$ as free parameters.} Our joint fitting result, presented in Figure~\ref{fig:MCMC quasar model}, indicates that the required quasar Ly$\alpha$ luminosity under the above assumption should be $\log [L_{q,\alpha}/({\rm erg\, s^{-1}})]=\vallogLq$. The best-fit value is even brighter than some ultraluminous quasars usually targeted to search for enormous nebulae (e.g., $\sim 10^{43}$ -- $\lesssim10^{45}{\rm erg\, s^{-1}}$ in \citealt{Cai2018}). Such a high \lya\ luminosity per quasar makes the quasar-dominated model unlikely to work. 

Our modeling result appears to be inconsistent with the quasar-dominated model in \citet{Croft2018}. In their model, the \lya\ SB profile on scales above $\sim 1\hinvMpc$ is well reproduced (see their Fig.10). \lya\ emission in their model is presented as \lya\ SB as a function of gas density and distance from the quasar, while the total \lya\ luminosity per quasar is not given. The luminosity, however, can be dominated on scales $\lesssim 1\hinvMpc$, which is not shown in their figure. Fortunately, panel (b) in their Figure 8 (``Model Q'') enables an estimation of the mean quasar \lya\ luminosity (R. Croft, private communication). With a mean \lya\ SB $\langle \mu_\alpha\rangle=7.0\times 10^{-22} {\rm erg\, s^{-1}cm^{-2}\textup{\AA}^{-1}arcsec^{-2}}$ (their section 5.1) from the slice with thickness of $40\hinvMpc$ (corresponding to observed spread of $\sim 29$\AA\ in \lya\ emission) and side length $400\hinvMpc$ ($\sim 2.04\times 10^4 {\rm arcsec}$ at $z\sim 2.5$), we obtain the total \lya\ luminosity in the slice to be $\sim 4.0\times 10^{47}{\rm erg\, s^{-1}}$. As there are about 100 quasars in the slice, the average \lya\ luminosity in ``Model Q'' of \citet{Croft2018} is $\sim 4.0\times 10^{45}{\rm erg\, s^{-1}}$, which agrees well with our result here.

\bigskip

In conclusion, the modeling results from our analytical models rule out the quasar-dominated scenario. 
For the galaxy-dominated scenario, however, both our measurement and that in \citet{Croft2018} imply that the detected Ly$\alpha$ signals cannot be explained simply by emission from currently observed LAEs. 
There must be additional Ly$\alpha$ emitting sources other than theses LAEs. We will explore the possibilities in Section~\ref{sec:discussion} after presenting the \lya\ forest-\lya\ emission cross-correlation results in Section~\ref{sec:forest correlation}.


\section{Ly$\alpha$ forest-Ly$\alpha$ emission cross-correlation}\label{sec:forest correlation}

Ly$\alpha$ forest, as a probe of the cosmic density field, can be used as an alternative tracer, more space-filling than quasars, to detect diffuse Ly$\alpha$ emission on cosmological scales. The \lya\ forest-\lya\ emission cross-correlation can provide additional information for understanding the origin of the \lya\ emission.

Following \citet{Croft2018}, we measure the Ly$\alpha$ forest-Ly$\alpha$ emission cross-correlation in a way similar to quasar-\lya\ emission cross-correlation:
\begin{equation}
\xi_{f\alpha }(r,\mu)=\frac{1}{\sum_{i=1}^{N(\vec{r})} w_{r i,\alpha} w_{r i, f}} \sum_{i=1}^{N(\vec{r})} w_{r i,\alpha} w_{r i, f} \Delta_{\mu, r i} \delta_{f, {ri}},
\end{equation}
where $N(\vec{r})$ is the number of Ly$\alpha$ forest-Ly$\alpha$ emission pixel pairs within the bin centered at the separation $\vec{r}=(r,\mu)$. $\Delta_{\mu, ri}$ is the fluctuation of Ly$\alpha$ emission SB (from the residual LRG spectra) for the $i$-th pixel pair in this bin, and $\delta_{f,ri}$ is the flux-transmission field of Ly$\alpha$ forest in the quasar spectra. The weights $w_{ri,\alpha}$ of \lya\ emission pixels are the same as in Equation~(\ref{eq:2dxcf}), and the weights for \lya\ forest pixels $w_{ri,f}=1/\sigma_{ri,f}^2$, where $\sigma_{ri,f}^2$ is the pixel variance due to instrumental noise and large scale structure, with the latter accounting for the intrinsic variance of the flux-transmission field.

Likewise, we can decompose the 2D \lya\ forest-\lya\ emission cross-correlation into the monopole, quadrupole and hexadecapole moments. To avoid spurious correlation induced by same-half-plate pixel pairs,  we only use pixel pairs residing on different half-plates and reject signals within $|r_{\|,{\rm cut}}|=4$cMpc, as discussed in Appendix~\ref{sec:half plate contamination}. Similar to what we do with the quasar-\lya\ emission cross-correlation, we define the modified multipoles of the \lya\ forest-\lya\ emission cross-correlation to be
\begin{equation}
\begin{aligned}
    \hat{\xi}_{f\alpha,\ell}(s) = \frac{2\ell+1}{2} 
      & \left( \int_{-1}^{-\mu_{\min}}\xi_{f\alpha}(s,\mu) \mathcal{L}_\ell(\mu)d\mu\right. \\
      + & \left. \int_{\mu_{\min}}^{1}\xi_{f\alpha}(s,\mu) \mathcal{L}_\ell(\mu)d\mu \right),
\end{aligned}
\end{equation}
where $\mu_{\min}=|r_{\|,{\rm cut}}|/s$. Like Equation~(\ref{eq:R_lk}), the original and modified multipoles are connected through $\hat{\xi}_{f\alpha} = \mathbf{R^\prime} \xi_{f\alpha}$,  where the element of the transformation matrix $\mathbf{R^\prime}$ takes the form of
\begin{equation}
\begin{aligned}
    R^{\prime}_{\ell k}=\frac{2\ell+1}{2} & \left(\int_{-1}^{-\mu_{\min}}\mathcal{L}_\ell(\mu)\mathcal{L}_k(\mu)d\mu\right. \\
    +  & \left. \int^{1}_{\mu_{\min}}\mathcal{L}_\ell(\mu)\mathcal{L}_k(\mu)d\mu \right),
\end{aligned}
\end{equation}
with $\ell,k=0$, 2, and 4.

The analytical model for the \lya\ forest-\lya\ emission cross-correlation is similar to the one for the quasar-\lya\ emission cross-correlation, and we only need to replace $b_q$ and $\beta_q$ in equations~(\ref{eq:model galaxy multipoles}) and (\ref{eq:model galaxy f}) with $b_f$ and $\beta_f$, respectively. Here $b_f$ is the \lya\ forest transmission bias, evolving with redshift as $b_{f}(z)=b_{f}(z_{\mathrm{ref}})[(1+z)/(1+z_{\mathrm{ref}})]^{\gamma_{\alpha}}$ with $\gamma_\alpha=2.9$, and $\beta_f$ is the redshift distortion parameter for Ly$\alpha$ forest, $\beta_f = f b_{\eta}/b_f$, where $f$ is the linear growth rate of
structure and $b_{\eta }$ is the velocity bias of Ly$\alpha$ forest \citep[e.g.,][]{Seljak2012,Blomqvist2019}. We fix $b_\eta=-0.225$ and $\beta_f=1.95$ at a reference redshift of $z_{\rm ref}=2.34$ according to the quasar-Ly$\alpha$ forest cross-correlation result in \citet{Bourboux2020}, yielding $b_f=-0.119$ at $z=2.41$.

\begin{figure*}
    \centering
    \includegraphics[width=\textwidth]{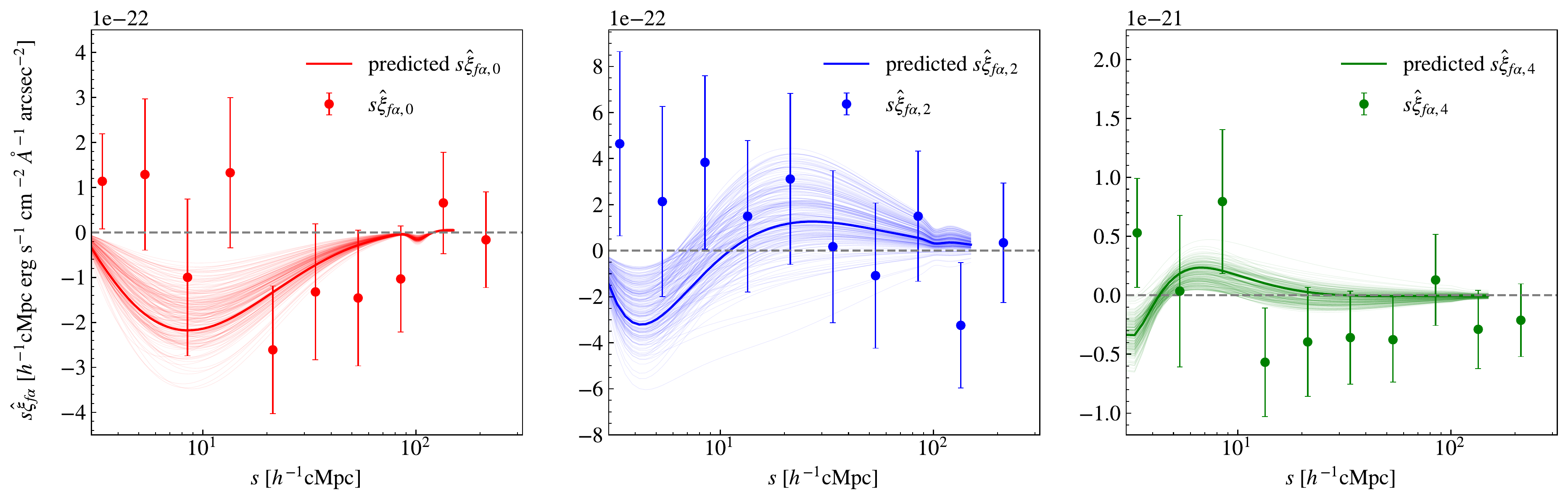}
    \caption{Modified monopole, quadrupole and hexadecapole of the Ly$\alpha$ forest-Ly$\alpha$ emission cross-correlation as a function of the Ly$\alpha$ forest-Ly$\alpha$ emission pixel pair separation. The data points are the measurements, and the solid curves are the predictions using parameters in literature to describe the \lya\ forest and parameters $\langle\mu_\alpha\rangle$ and $\beta_\alpha$ derived from fits to the quasar-\lya\ emission cross-correlation under the galaxy-dominated scenario. The various solid curves are the predicted modified multipoles from randomly drawing $\langle\mu_\alpha\rangle$ and $\beta_\alpha$ from their posterior probability distribution, with the thickest ones from adopting the best-fit parameters.
    }
    \label{fig:xi fa}
\end{figure*}

Given the small transmission bias $b_f$ of Ly$\alpha$ forest, the expected \lya\ forest-\lya\ emission cross-correlation level at $\sim 10h^{-1}$cMpc is $\sim 5\%$ of the quasar-Ly$\alpha$ emission cross-correlation. The subsequent low signal-to-noise ratio would lead to weak parameter constraints from fitting the Ly$\alpha$ forest-Ly$\alpha$ emission cross-correlation measurements. Instead we choose to compare the measurements with the predictions from the model adopting the best-fit parameters, $\beta_\alpha$ and $\langle\mu_\alpha\rangle$, from modeling the quasar-\lya\ emission correlation (Section~\ref{sec:model galaxy}). Such a consistency check is shown in Figure~\ref{fig:xi fa}.

\add{The multipole measurements in Figure~\ref{fig:xi fa} indicate that there is no significant detection of the \lya\ forest-\lya\ emission cross-correlation. Quantitatively, a line of zero amplitude would lead to $\chi^2=19.8$ for a total of 21 data points of the monopole, quadrupole, and hexadecapole in the range of $4\hinvMpc< s <100\hinvMpc$.
On the other hand, with the large uncertainties in the data, our model predictions also appear to be consistent with the measurements. The predictions from the bestfit model (solid curves) give a value of $\chi^2=29.5$ for the above 21 data points, within $\sim$1.3$\sigma$ of the expected mean $\chi^2$ value.
}
We note that the monopole is consistent with that in \citet{Croft2018}, as long as the uncertainties are taken into account (see their Fig.11). Our model has a much lower amplitude than their galaxy-dominated model (model G), leading to a closer match to the data. This is a manifestation of the lower $\langle \mu_\alpha \rangle$ value inferred from our quasar-\lya\ emission cross-correlation measurements.

\section{Discussion: possible Ly$\alpha$ sources}\label{sec:discussion}

Our quasar-\lya\ emission cross-correlation measurements can be explained by a model with \lya\ emission associated with star-forming galaxies (\S~\ref{sec:xcf}), and the \lya\ forest-\lya\ emission cross-correlation measurements are also consistent with such an explanation (\S~\ref{sec:forest correlation}). The model, however, does not provide details on the relation between \lya\ emission and galaxies, which we explore in this section.

As shown in Figure~\ref{fig:luminosity density compare}, the measured Ly$\alpha$ luminosity density, $\rho_{\rm Ly\alpha}=\valrholya\times 10^{40}$ erg s$^{-1}$ cMpc$^{-3}$, computed from our best-fit $\langle\mu_\alpha\rangle$ under the galaxy-dominated case. This iceberg of \lya\ emission can hardly be accounted for by \lya\ emission from LAEs based on observed \lya\ LFs, as shown in Figure~\ref{fig:luminosity density compare} with \lya\ luminosity densities obtained from integrating the \lya\ LF of LAEs down to a low luminosity. For example, the value of $\rho_{\rm Ly\alpha}$ calculated by integrating the LAE LF at $z=2.5\pm 0.1$ in \citet{Sobral2018} down to $1.75\times10^{41} {\rm erg ~s^{-1}}$ is $7.4^{+0.8}_{-0.7}\times 10^{39} {\rm erg ~s^{-1} ~cMpc^{-3}}$, only $\sim 12\%$ of our estimate. That is, \lya\ emission formally detected from LAEs is only the tip of the iceberg. 

Conversely, if we assume that all the Ly$\alpha$ photons detected in our work are produced by star formation activities and neglect any dust effect on \lya\ emission, the implied SFRD $\rho_{\rm SFR}$ approximates the lower bound of the dust-corrected cosmic $\rho_{\rm SFRD}$ determined by UV and IR observations (see Figure~\ref{fig:SFRD compare}).

There have to be some other sources responsible for the excessive Ly$\alpha$ emission. In this section, we explore two possible sources based on previous observations and models:
\lya\ emission within an aperture centered on star-forming galaxies, including LAEs and Lyman break galaxies (LBGs), with a typical aperture of $2\arcsec$ in diameter in most NB surveys; \lya\ emission outside the aperture usually missed for individual galaxies in NB surveys, commonly called extended or diffuse \lya\ halos. We name the two components as inner and outer part of \lya\ emission, respectively. For the outer, diffuse \lya\ halo component, we do not intend to discuss its origin here \citep[e.g.,][]{Zheng2011b,Lake2015} but adopt an observation-motivated empirical model to estimate its contribution.

We argue that almost all star-forming galaxies produce Ly$\alpha$ emission, and actually, significant emission may be originated from their halos. This should contribute to the bulk of faint diffuse Ly$\alpha$ emission in the Universe, as detected in this work. 

\subsection{Inner Part of Ly$\alpha$ Emission for UV-selected Star-forming Galaxies }\label{sec:rho by REW}

A large portion of LBGs exhibit Ly$\alpha$ emission, though their rest-frame equivalent width (REW) might not satisfy the criteria for LAE selections \citep{Shapley2003,Vieuville2020} if measured with a typical aperture of $2\arcsec$ in diameter in NB surveys. It is also detected in deep stacks of luminous and massive LBGs \citep{Steidel2011} and in individual UV-selected galaxies in recent MUSE eXtremely Deep Field (MXDF) observations \citep{Kusakabe2022}.  

\citet{Dijkstra2012} reported the Ly$\alpha$ REW distribution of $\sim 800$ $z\sim 3$ LBGs spectroscopically observed by \citet{Shapley2003} with $1.4\arcsec$ slits, which can be described well by an exponential function. This sample includes both \lya\ emission ($\REW>0$~\AA) and \lya\ absorption ($\REW<0$~\AA) within the central aperture. Combined with this empirical model of \lya\ $\REW$ distribution for star-forming galaxies, we perform integration over the UV LF  
to obtain the corresponding \lya\ luminosity density,
\add{
\begin{equation}\label{eq:rho_lya inner}
\begin{aligned}
\rho^{\rm inner}_{\rm Ly\alpha} = & \int_{M_{\rm UV,min}}^{M_{\rm UV,max}}  [ \langle L_\alpha(M_\UV)\rangle \Phi^e_\UV(M_\UV) 
\\
&+ \langle L^{\rm Abs}_\alpha(M_{\rm UV})\rangle \Phi^a_{\rm UV}(M_{\rm UV}) ] dM_{\rm UV}
\end{aligned}
\end{equation}}
where $\langle L_\alpha(M_\UV)\rangle$ is the mean \lya\  luminosity within the aperture of the $\REW >0$~\AA\ population at a given UV luminosity \add{and $\langle L^{\rm Abs}_\alpha(M_{\rm UV})\rangle$ is the absorption of the $\REW <0$~\AA\ population making a negative contribution}. The function $\Phi^e_\UV$ is the UV LF for the $\REW>0$~\AA\ population, which is the overall UV LF $\Phi_\UV$ multiplied by the (UV luminosity-dependent) fraction of such a population, \add{and $\Phi^a_\UV$ for the $\REW<0$~\AA\ population likewise}. More details on the calculations in our adopted model are presented in Appendix~\ref{sec: Lya model}.

We select five observed UV LFs around $z\approx 2.4$ from the literature (Table~\ref{tab:UV LF}), and calculate the corresponding Ly$\alpha$ luminosity densities, which are shown in Table~\ref{tab:rho Lya}. 

\add{
We note that the distribution of \lya\ REW within the central aperture is mainly determined by three factors: the intrinsic REW from photoionization and recombination in the \ion{H}{2} region of star-forming galaxies, the dust extinction, and the scattering-induced escape fraction. The empirically modelled \lya\ REW distribution in \citet{Dijkstra2012} we adopt reflects the combination of the three factors.
}


\begin{table*}
\centering
\caption{A compilation of the derived Schechter function parameters for the galaxy UV LFs adopted in this work.}
\begin{threeparttable}
\begin{tabular}{cccccc}
\hline
\hline
Source & $z$ & $\lambda_{\rm UV}$\tnote{1} (\AA) & $M^*$ &
$\Phi^*(10^{-3} {\rm cMpc}^{-3})$& $\alpha$ \\
\hline
\citet{Reddy2009} & 2.3 & 1700 & $-20.70 \pm 0.11$ & $2.75 \pm 0.54$ & $-1.73 \pm 0.07$ \\
\citet{Sawicki2012} & 2.2 & 1700 & $-21.00\pm0.50$ & $2.74\pm 0.24$ &  $-1.47\pm 0.24$  \\
\citet{Parsa2016} & 2.25& 1700 &  $-19.99 \pm 0.08$ & $6.20 \pm 0.77$ & $-1.31 \pm 0.04$ \\
\hline
\citet{Bouwens2015}  & - & 1600 & $-20.89+0.12z$ & $0.48 \times 10^{-0.19(z-6)}  $ & $-1.85 - 0.09(z - 6)$ \\
extrapolation\tnote{2} & 2.4 & & -20.60 & 2.3 & -1.53 \\ 
    \hline
\citet{Parsa2016} & - & 1700 & $\frac{-35.4 (1+z)^{0.524}} {1+(1+z)^{0.678}}$ & $ -0.36 z + 2.8 $ & $ -0.106 z -1.187$ \\
extrapolation\tnote{3} & 2.4 & & -20.41 & 1.9 & -1.44 \\
\hline
\hline
\end{tabular}
\begin{tablenotes}[flushleft]
    \footnotesize
    \item[1] Rest-frame UV wavelength where the UV LF is measured. Note that $\lambda_{\rm UV}$ for \cite{Bouwens2015} is 1600 \AA, while the empirical model in \citet{Dijkstra2012} as summarized in Appendix \ref{sec:Dijkstra model} adopts 1700 \AA. We just assume that UV LFs are not sensitive to such a subtle difference in $\lambda_{\rm UV}$.
    
    \item[2] Extrapolation of the  Schechter parameters of the UV LF to $z=2.4$ adopting the best-fitting formula in \citet{Bouwens2015} for the redshift evolution.
     
    \item[3] Extrapolation to $z=2.4$, based on the simple parametric fits to published Schechter parameters in \citet{Parsa2016}. Note that this fitting is meant to illustrate the overall evolutionary trend, but not to indicate a best estimate of true parameter evolution. 
\end{tablenotes}
\end{threeparttable}
\label{tab:UV LF}
\end{table*}

\subsection{Outer Part of Ly$\alpha$ Emission from Galaxy Halos}\label{sec:rho by halo}

\add{As discussed before}, many previous works have reported detections of extended Ly$\alpha$ emission around high-redshift galaxies, either by discoveries of Ly$\alpha$ halos/blobs around bright individual star-forming galaxies through ultradeep exposures \citep{Steidel2000,Matsuda2004,Matsuda2011,Wisotzki2016,Leclercq2017,Kusakabe2022}, or by employing stacking analyses on large samples \citep{Steidel2011,Matsuda2012,Momose2014,Momose2016,Xue2017}.  Most extended Ly$\alpha$-emitting halos are discovered around LAEs \citep{Wisotzki2016,Leclercq2017}; \add{they are also prevalent around non-LAEs, e.g.,  UV-selected galaxies, due to a significant amount of cool/warm gas in their CGM \citep{Steidel2011,Kusakabe2022}. }

The cumulative fraction of the large-aperture Ly$\alpha$ flux, shown in Fig.10 of \citet{Steidel2011}, 
indicates that a 2 arcsec aperture adopted by typical deep narrow/medium-band LAE surveys 
could miss $\sim$50\%  Ly$\alpha$ emission for  
LBGs with net (positive) Ly$\alpha$ emission. Thus Equation~(\ref{eq:rho_lya inner}) could underestimate the total Ly$\alpha$ flux from ${\REW}>0$~\AA\ galaxies roughly by a factor of 2. For galaxies whose inner parts present net Ly$\alpha$ absorption, the existence of extended Ly$\alpha$ halos has been strongly confirmed by the sample with Ly$\alpha$ ${\REW}<0$~\AA\ in \cite{Steidel2011}, whose radial SB profile outside 10 kpc is qualitatively similar to that of the non-LAE sub-samples.  

Given the above observational results, we adopt the reasonable model that all star-forming galaxies, whether showing \lya\ emission or absorption within the central aperture, have \lya\ emitting halos.
Based on the strong anti-correlation between Ly$\alpha$ luminosities of Ly$\alpha$ halos and the corresponding UV magnitudes reported in \citet{Leclercq2017}, we assume that the Ly$\alpha$ luminosity from halos of galaxies with ${\REW}<0$~\AA\ depends on $M_{\rm UV}$ only. We further assume that it is equal to the inner part originated from the ${\REW}>0$~\AA\ galaxy population at a given $M_\UV$ \citep{Steidel2011}. Therefore we express the total contribution to the \lya\ luminosity density from the outer part as 
\begin{equation} \label{eq:rho_lya outer}
    \rho^{\rm outer}_{\rm Ly\alpha} = \int_{M_{\rm UV,min}}^{M_{\rm UV,max}} \langle L_\alpha(M_\UV)\rangle \Phi_\UV(M_\UV) dM_\UV,
\end{equation}
where $\Phi_\UV$ denotes the UV LF for the entire population (See Appendix \ref{sec: Lya model}). 

\bigskip

\begin{figure}[htbp]
    \includegraphics[width=.5\textwidth]{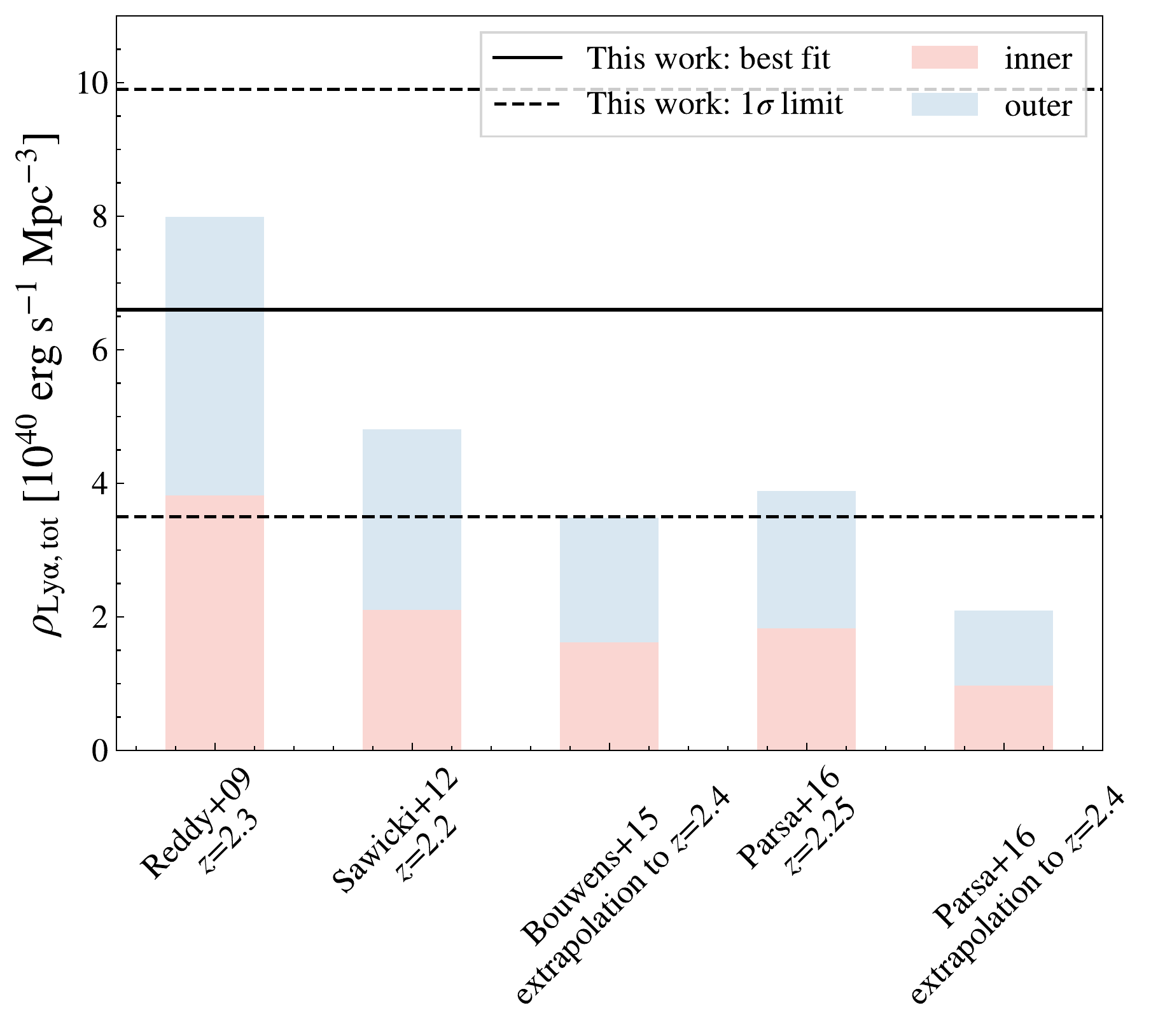}
    \caption{Ly$\alpha$ luminosity density computed in our model by integrating different observed UV LFs. Different colors denote the inner and outer Ly$\alpha$ parts, as described in Table~\ref{tab:rho Lya} in details. The model Ly$\alpha$ luminosity densities are compared with that inferred from our quasar-\lya\ emission cross-correlation measurements, $6.6\times 10^{40}$ erg s$^{-1}$ cMpc$^{-3}$ (solid) with 1$\sigma$ errorbars (dashed) of $\pm 3.2\times 10^{40}$ erg s$^{-1}$ cMpc$^{-3}$.
    }
    \label{fig:rho Lya estimate}
\end{figure}

Clearly, the total \lya\ luminosity density should be $\rho_{\rm Ly\alpha}^{\rm tot}=\rho_{\rm Ly\alpha}^{\rm inner}+\rho_{\rm Ly\alpha}^{\rm outer}$. Note that the total \lya\ luminosity density estimated from the model is just a lower limit as discussed in Appendix \ref{sec: Lya model}, since we (1) adopt a constant scaling factor for the \cite{Dijkstra2012} empirical model and (2) use this empirical model that is designed for the $\REW >0$~\AA\ population to describe the $\REW <0$~\AA\ one. A brief summary of the estimated Ly$\alpha$ luminosity density is listed in Table~\ref{tab:rho Lya} and Figure~\ref{fig:rho Lya estimate}. Revealed by Figure~\ref{fig:rho Lya estimate}, the total Ly$\alpha$ luminosity density derived from our model is consistent with our detection within 1$\sigma$ (or $\sim 1.3\sigma$ when using the $z=2.4$ UV LF in \citealt{Parsa2016}). We argue that, star-forming galaxies, which contain the inner part of \lya\ emission that can be captured by the aperture photometry in deep NB surveys and the outer part of \lya\ emission from their halos, usually outside the aperture, could produce sufficient Ly$\alpha$ emission to explain our detection from the quasar-\lya\ emission cross-correlation measurement.

Our derived $\rho_{\rm Ly\alpha}$ is higher than the result of \cite{Wisotzki2018}, who use MUSE observations of extended \lya\ emission from LAEs to infer a nearly 100\% sky coverage of \lya\ emision. The LAE sample they use are selected from the Hubble Deep Field South (HDFS) and the Hubble Ultra Deep Field (HUDF), a subset of LAEs whose \lya\ LFs has been analyzed in \citet{Drake2017} and \citet{Drake2017_south} (though the sample in \cite{Wisotzki2018} contains a few additional LAEs). As shown in Figure~\ref{fig:luminosity density compare}, $\rho_{\rm Ly\alpha}$ estimated in \cite{Drake2017} is lower than ours, too. Our result implies that \cite{Wisotzki2018} may underestimate the \lya\ sky coverage at a given SB level when simply focusing on LAEs and ignoring the diffuse \lya\ emission from faint UV-selected galaxies. 

\add{As shown in Figure \ref{fig:rho Lya estimate}, about half of the detected \lya\ photons come from the inner part of galaxies. By assuming that they all stem from star formation activities, we estimate the escape fraction $f_{\rm esc}$ for these \lya\ photons to be roughly $0.21_{-0.11}^{+0.21}$, where the cosmic intrinsic \lya\ luminosity density due to star formation is calculated based on the cosmic SFRD shown in Figure \ref{fig:SFRD compare}, yielding \addb{$1.44^{+10.1}_{-6.1} \times 10^{41} {\rm erg\, s^{-1} Mpc^{-3}}$}. While the estimated $f_{\rm esc}$ appears consistent with previous work within $1\sigma$ uncertainties (e.g., $\sim 10\%$ in \citealt{Chiang2019}), we emphasize that the galaxy population involved in our modelling is different
from LAEs in typical NB surveys. We include galaxies with low \lya\ REW usually not identified as LAEs, which boost our estimate for $f_{\rm esc}$ compared with LAE-derived ones.}

\begin{table}
\centering
    \caption{Model Ly$\alpha$ luminosity density $\rho_{\rm Ly\alpha}$ by integrating UV LFs (from $M_{\rm UV,min}=-24$ to $M_{\rm UV,max}=-12$) based on Schechter functions from various sources as in Table~\ref{tab:UV LF}. See Section~\ref{sec:rho by REW} and \ref{sec:rho by halo} for more details.}
    \resizebox{\columnwidth}{!}{
    \begin{threeparttable}[t]
    \begin{tabular}{ccccc}
    \hline
    \hline
    Source & $z$ & \multicolumn{3}{c}{$\rho_{\rm Ly\alpha}$ ($10^{40}$ erg s$^{-1}$ cMpc$^{-3}$)} \\
    \hline
    & & inner\tnote{1} & outer\tnote{2} & total\tnote{3}  \\
    \hline
    \citet{Reddy2009} & 2.3 & 3.82 & 4.17  & 8.00\\
    \citet{Sawicki2012} & 2.2 & 2.10 & 2.71   &  4.81  \\
    \citet{Parsa2016} & 2.25& 1.83 & 2.05  & 3.88\\
    \hline
    \citet{Bouwens2015}  & \multirow{2}{*}{2.4} &  \multirow{2}{*}{1.61} & \multirow{2}{*}{1.87} &  \multirow{2}{*}{3.49}\\
    extrapolation\tnote{4} &  \\
    \citet{Parsa2016} & \multirow{2}{*}{2.4} &  \multirow{2}{*}{0.97} & \multirow{2}{*}{1.12} &   \multirow{2}{*}{2.09}  \\
    extrapolation & \\
    \hline
    \hline
    \end{tabular}
    \begin{tablenotes}[flushleft]
    
        \footnotesize
        \item[1] Ly$\alpha$ luminosity density from emission that would be captured within an aperture of $2\arcsec$ in diameter, computed from Equation~(\ref{eq:rho_lya inner}). Galaxies with Ly$\alpha$ ${\REW}>0$~\AA\ contribute a positive part and the Ly$\alpha$ ${\REW}<0$~\AA\ population contribute a negative one.
         
        \item[2]  Ly$\alpha$ luminosity density from emission outside the 2$\arcsec$ aperture for all galaxies, i.e., the diffuse \lya\ halo component, computed from Equation~(\ref{eq:rho_lya outer}). At a given UV luminosity, we assume that the populations with central $\REW >0$~\AA\ and $\REW<0$~\AA\ have the same diffuse halo \lya\ luminosity, which is set to be the same as that from the inner part of the $\REW >0$~\AA\ population in our model based on the results in \citet{Steidel2011}.
        
        \item[3] Total Ly$\alpha$ luminosity density contributed by the three components discussed above.

        \item[4] Same as in Table~\ref{tab:UV LF}.
       
    \end{tablenotes}
    
    \end{threeparttable}
    }
    
    \label{tab:rho Lya}
\end{table}

\section{Summary and conclusion}

In this work, we have performed a cross-correlation analysis of the SDSS BOSS/eBOSS LRG residual spectra at wavelengths $\lambda$ = 3647--5471\AA\ and DR16 quasars at a redshift range of $2<z<3.5$. This enables a measurements of the cross-correlation between quasar position and \lya\ emission intensity (embedded in the residual LRG spectra) at a median redshift $z\sim 2.4$.
The Ly$\alpha$ SB profile around quasars is obtained by projecting our cross-correlation results into a pseudo-narrow band, and the truncated forms of the monopole, quadrupole and hexadecapole of the quasar-Ly$\alpha$ emission cross-correlation are computed by discarding small-scale signals within $r_{\perp}< 4 h^{-1}{\rm cMpc}$. 

Our work improves upon that in \citet{Croft2018} by making use of the final SDSS-IV release of LRG spectra and quasar catalog. While our \lya\ SB profile measurements are consistent with that in \citet{Croft2018}, our inferred large-scale clustering amplitude is about 2.2 times lower. \add{Although the absolute uncertainty in our work is about 25\% lower, the lower clustering amplitude leads to a larger fractional uncertainty. This is a reflection of our more rigorous treatment to possible contaminated fibers and our exclusion of the small-scale signals in modelling the multipoles.
}
With this lower amplitude, our measured \lya\ forest-\lya\ emission cross-correlation can also be consistently explained.

Like \citet{Croft2018}, on sub-Mpc scales the obtained \lya\ SB forms a natural extrapolation of that observed from the luminous Ly$\alpha$ blobs on smaller scales \citep{Borisova2016,Cai2019}. Unlike \citet{Croft2018}, we find that the amplitudes of the large-scale \lya\ SB and quasar-\lya\ emission cross-correlation cannot result from the \lya\ emission around quasars, as this would require the average \lya\ luminosity of quasars to be about two orders of magnitude higher than observed given their rather low number density.

To figure out the most possible sources that contribute to the detected Ly$\alpha$ signals, we construct a simple analytical model, which combines the SB profile and multipole measurements. The inferred Ly$\alpha$ luminosity density, $\valrholya\times 10^{40} {\rm erg\, s^{-1} cMpc^{-3}}$, is much higher than those from integrating the \lya\ LFs of LAEs. \add{We fix the luminosity weighted bias of galaxies $b_\alpha$ to be 3 in our modelling, which turns out to be a good estimate. But bear in mind that the luminosity density scales with 3/$b_\alpha$ if $b_\alpha$ deviates from that value.} Our model rules out the possibility that the diffuse emission is due to reprocessed energy from the quasars themselves, and support the hypothesis that star-forming galaxies clustered around are responsible for the detected signal. \add{
For the Ly$\alpha$ forest-Ly$\alpha$ emission cross-correlation, the prediction from our model matches the measurement, although the current measurement is consistent with a null detection given the low signal-to-noise ratio.}
We argue that most star-forming galaxies exhibit Ly$\alpha$ emission. These include galaxy populations with either \lya\ emission or \lya\ absorption at the center, while both populations have diffuse \lya\ emitting halos, which are usually missed in individual LAEs from deep narrow-band  surveys. Our estimates based on the empirical model of \cite{Dijkstra2012} and the observed UV LFs of star-forming galaxies are able to match the \lya\ luminosity density inferred from our cross-correlation measurements. The picture is supported by stacked analysis from NB surveys \citep[e.g.,][]{Steidel2011} and by the IFU observations of \lya\ emission associated with UV-selected galaxies \citep[e.g.,][]{Kusakabe2022}.

Our work shows an enormous promise of Ly$\alpha$ intensity mapping as a probe of large scale structure. One can also utilize this technique to explore the intensity of other spectral lines, once a larger data set is provided.  The next-generation cosmological spectroscopic survey, the ongoing Dark Energy Spectroscopic Instrument (DESI; \citealt{DESI16}), will enlarge the galaxy/quasar survey volume at least by an order of magnitude compared to SDSS BOSS/eBOSS. We expect the intensity mapping technique carried out in DESI will bring us new insights into the Universe. Deep surveys of \lya\ emission around star-forming galaxies, especially the UV-selected population \citep[e.g.,][]{Kusakabe2022} will shed light on the intensity mapping measurements and provide inputs for building the corresponding model.  Moreover, more realistic modelling of physical processes such as radiative transfer and quasar proximity effect should be considered to advance our understanding of the Ly$\alpha$ emission iceberg in the Universe.

\acknowledgments

We thank Kyle Dawson, Rupert Croft, and Coast Zhang for useful discussions. X.L. and Z.C. are supported by the National Key R\&D Program of China (grant
No.2018YFA0404503) and the National Science Foundation
of China (grant No. 12073014). Z.Z. is supported by NSF grant AST-2007499.

Funding for the Sloan Digital Sky 
Survey IV has been provided by the 
Alfred P. Sloan Foundation, the U.S. 
Department of Energy Office of 
Science, and the Participating 
Institutions. 

SDSS-IV acknowledges support and 
resources from the Center for High 
Performance Computing  at the 
University of Utah. The SDSS 
website is www.sdss.org.

SDSS-IV is managed by the 
Astrophysical Research Consortium 
for the Participating Institutions 
of the SDSS Collaboration including 
the Brazilian Participation Group, 
the Carnegie Institution for Science, 
Carnegie Mellon University, Center for 
Astrophysics | Harvard \& 
Smithsonian, the Chilean Participation 
Group, the French Participation Group, 
Instituto de Astrof\'isica de 
Canarias, The Johns Hopkins 
University, Kavli Institute for the 
Physics and Mathematics of the 
Universe (IPMU) / University of 
Tokyo, the Korean Participation Group, 
Lawrence Berkeley National Laboratory, 
Leibniz Institut f\"ur Astrophysik 
Potsdam (AIP),  Max-Planck-Institut 
f\"ur Astronomie (MPIA Heidelberg), 
Max-Planck-Institut f\"ur 
Astrophysik (MPA Garching), 
Max-Planck-Institut f\"ur 
Extraterrestrische Physik (MPE), 
National Astronomical Observatories of 
China, New Mexico State University, 
New York University, University of 
Notre Dame, Observat\'ario 
Nacional / MCTI, The Ohio State 
University, Pennsylvania State 
University, Shanghai 
Astronomical Observatory, United 
Kingdom Participation Group, 
Universidad Nacional Aut\'onoma 
de M\'exico, University of Arizona, 
University of Colorado Boulder, 
University of Oxford, University of 
Portsmouth, University of Utah, 
University of Virginia, University 
of Washington, University of 
Wisconsin, Vanderbilt University, 
and Yale University.


\appendix

\section{Correcting Measurement Systematics}\label{sec:systematics}

Dealing with possible contamination is a difficult problem in all intensity mapping (IM) experiments. Since the expected signals in our measurement have gone beyond the detection capability of any current instruments, it is crucial to remove possible systematics. In this section we discuss three main sources of potential contamination: cross-talk effect among spectra in adjacent fibers, correlation at $r_\|=0$ for pixel-pixel correlation, and spurious signal on larger scales, and then demonstrate that we have removed them carefully from our measurement.

\subsection{Quasar Stray Light Contamination}\label{sec:stray light systematics}
The BOSS/eBOSS spectrograph has 1,000 fibers per plate, which disperse light onto the same 4096-column CCD. Light from one fiber would possibly leak into the extraction aperture for another fiber, but the level of this light contamination is negligible in the SDSS data reduction pipeline. However, our intensity-mapping technique reaches far beyond the instrument capability ($\sim 10^{-17}\SBunitsA$), so this light contamination should be treated cautiously. When cross-correlating
quasar-LRG spectrum pixels, the cross-correlation between quasars and its leak into LRG spectra will lead to
a contamination 3-4 orders of magnitude higher than the
targetted Ly$\alpha$ signals, due to the bright and broad Ly$\alpha$
features of quasars. 

In \citet{Croft2016}, quasar stray light contamination is removed through discarding any quasar-LRG spectrum pixel pairs once on the CCD the quasar is within five fibers apart from the LRG, i.e., $\Delta{\rm fiber}< 5$. Moreover, \citet{Croft2018} reported that the remaining quasar stray light would still lead to contamination as a result of quasar clustering effect: an LRG fiber with $\Delta_{\rm fiber}\geq 5$ away from a quasar fiber may be contaminated by another quasar, and if not corrected for, the cross-correlation between \lya\ emission in the LRG fiber with the first quasar would have the quasar clustering signal imprinted.
They find that the quasar clustering effect would reach 50\% of the signal in \citet{Croft2016}
on scales of $|r_\perp| < 10 \hinvMpc$ and $|r_\| |< 10 \hinvMpc$. \citet{Croft2018} correct such an effect by generating a set of mock spectra, which contain quasar contaminating light only, and performing the same cross-correlation procedure to measure the intensity of clustering. Then the clustering signal from the mock is subtracted from their originally measured signals. 

\begin{figure}[htbp]
    \centering
    \includegraphics[width=.5\textwidth]{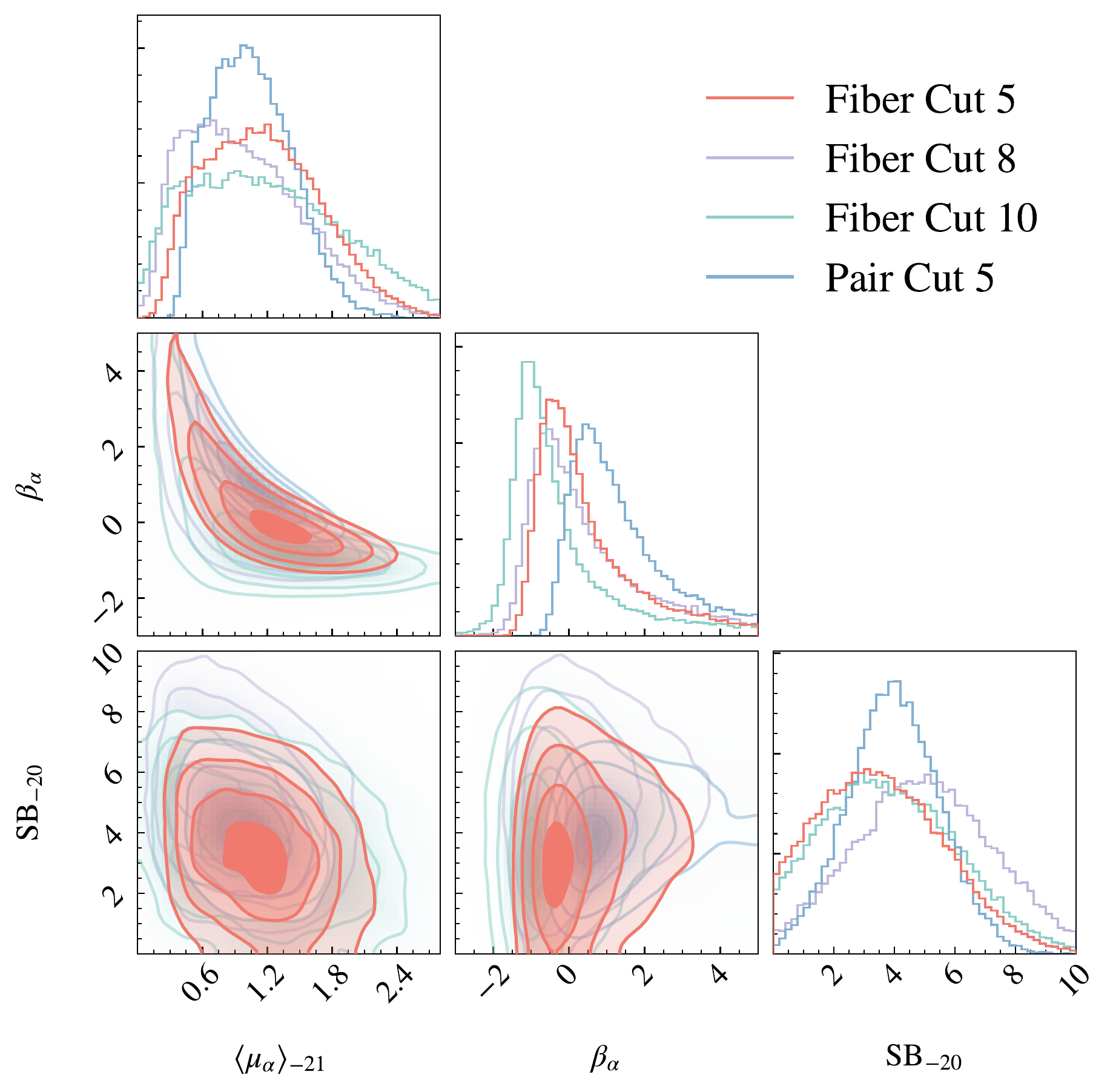}
    \caption{A test for the robustness of our fiducial sample selection, by performing joint fits to the multipoles and SB profiles of quasar-Ly$\alpha$ emission cross-correlation after different sample selections. The model parameters are the same as in Figure~\ref{fig:MCMC galaxy model}. \texttt{Fiber Cut 5} refers to the sample that any LRG fiber with $\Delta{\rm fiber}<5$ from a quasar fiber is excluded, which is the fiducial sample for our measurements.; \texttt{Fiber Cut 8} and \texttt{Fiber Cut 10} refers to $\Delta{\rm fiber}<8$ and $\Delta{\rm fiber}<10$, respectively. \texttt{Pair Cut 5} refers to the sample in which a quasar-LRG spectrum pixel pair is excluded if they satisfies $\Delta{\rm fiber}<5$, i.e. the method adopted in \citet{Croft2018}. 
    }
    \label{fig:Fiber Cut}
\end{figure}

The key of the algorithm in \citet{Croft2018} is to estimate the light leakage fraction so that cross-correlation of mock spectra can precisely reproduce the quasar clustering effect. The fraction measured in \citet{Croft2016} is no longer applicable to our sample spectra, however, due to the recent updates on DR16 optical spectra pipeline\footnote{\url{https://www.sdss.org/dr16/spectro/pipeline/}}. In our measurement, to be conservative, instead, we exclude any LRG fiber once it is within 5 fiber or less apart from a quasar fiber, and this fiducial sample selection will remove both quasar stray light contamination and quasar clustering effect simultaneously. We also repeat the algorithm introduced in \citet{Croft2018}, removing the quasasr clustering systematics by subtracting the cross-correlation pattern produced by mock spectra, and then measure the corresponding multipoles and SB profiles. To ensure the robustness of our fiducial sample selection, i.e., excluding all LRG fibers of $\Delta{\rm fiber}<5$, we perform the same fitting procedure mentioned in Section~\ref{sec:model} under the galaxy-dominated assumption for test cases with various sample selections. A comparison of results with differently selected samples is demonstrated in Figure~\ref{fig:Fiber Cut}. The result of $\Delta{\rm fiber}<5$ LRG exclusion are in fact consistent with that of $\Delta{\rm fiber}<8$ and $\Delta{\rm fiber}<10$ within $1\sigma$, implying that Four fiducial selection can remove the contamination well. It in general accords with the result of $\Delta$fiber$<5$ pair exclusion, i.e., the method used in \citet{Croft2018}, though there is a tiny offset in best-fit $\beta_\alpha$ and the uncertainties of the three parameters from the latter are smaller.

\subsection{$r_\|=0$ Correlation for Pixel-pixel Pairs}\label{sec:half plate contamination}

In measuring the \lya\ forest-\lya\ emission cross-correlation, we need to remove an artifact of correlation around $r_\|=0$, introduced by the spectra pipeline.

In BOSS/eBOSS, each half-plate has $\sim$500 fibers (450 science fibers and $\sim$ 40 sky fibers) with two spectrographs. The sky-subtraction for individual spectra is done indepently for each spectrograph. Poisson fluctuations in sky spectra will induce correlations in those spectra obtained with the same spectrograph at the same observed wavelength \citep{Bautista2017,Bourboux2019,Bourboux2020}. Namely, a positive correlation is expected for spectrum pixel pairs on the same half-plates at $\Delta\lambda_{\rm obs}=0$, leading to an excess correlation in $r_\| = 0$ bins. Furthermore, the continuum fitting procedure designed for Ly$\alpha$ forest transmission fields may smooth the excess correlation at $r_\| = 0$, extending it to larger $|r_\| |$. Therefore we reject Ly$\alpha$ forest-Ly$\alpha$ emission pixel pairs once they are observed on the same-half plate. 

\begin{figure}
    \centering
    \includegraphics[width=\textwidth]{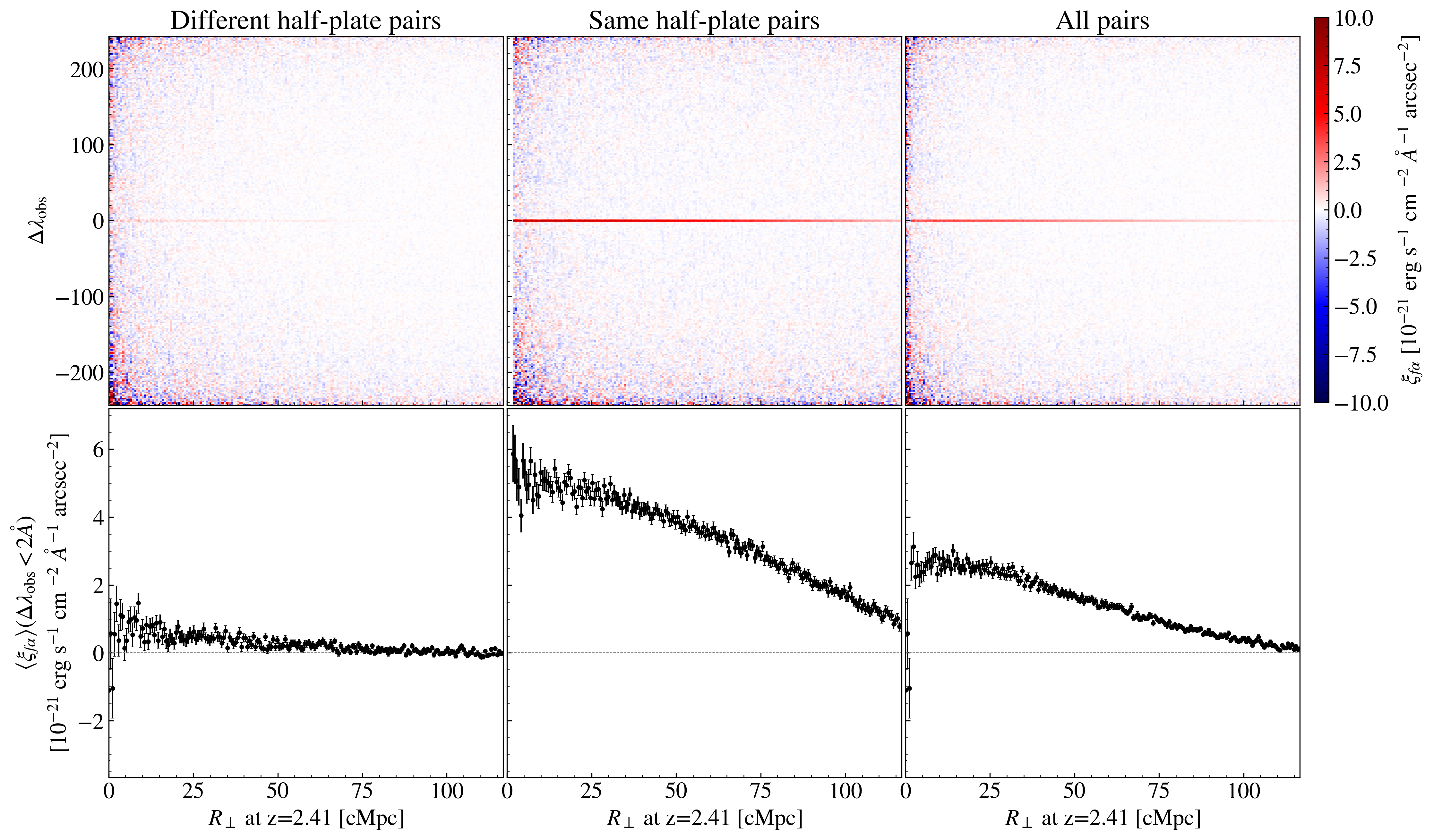}
    \caption{Ly$\alpha$ forest-Ly$\alpha$ emission pixel-pixel correlation as a function of observed wavelength separations $\Delta\lambda_{\rm obs}$ and transverse separations at a redshift of 2.41. \texttt{Different half-plate pairs} refers to the sample in which the selected pixel pairs reside on different half-plates; \texttt{Same half-plate pairs} refers to the sample in which all pairs are on the same half-plates.  \texttt{All pairs} refers to the sample without preference for the plates/fibers. The bottom panels show the correlations averaged within $2\textup{\AA}$ at $\Delta\lambda_{\rm obs}\sim 0$.
    }
    \label{fig:fa lambda cor}
\end{figure}

To evaluate how this same-spectrograph induced systematics would contaminate the signals and whether we have fully removed it, we perform a measurements of cross-correlation  between Ly$\alpha$ forest transmission pixels and Ly$\alpha$ emission pixels, as a function of their observed wavelength separations $\Delta\lambda_{\rm obs}$ and transverse separations $\Delta\theta$,
\begin{equation}
\xi_{f\alpha }(\Delta\lambda_{\rm obs},\Delta\theta)=\frac{1}{\sum_{i=1}^{N} w_{r i,\alpha} w_{r i, f}} \sum_{i=1}^{N} w_{r i,\alpha} w_{r i, f} \Delta_{\mu, r i} (\lambda_{\rm obs},\theta) \delta_{f, {ri}} (\lambda_{\rm obs} + \Delta\lambda_{\rm obs}, \theta + \Delta\theta),
\end{equation}
where $\Delta\theta$ can be easily converted to the transverse comoving separation at $z=2.41$ by $R_{\perp}=\Delta\theta \cdot D_{\rm C}(z=2.41)$.  The cross-correlation results of different-half-plate pixel pairs, same-half-plate pixel pairs and all pixel pairs without selection preference are shown in Figure~\ref{fig:fa lambda cor}. The contamination at $\lambda_{\rm obs}=0$ for same-half-plate pairs reaches several times $10^{-21}\SBunitsA$ (middle panel), even stronger than the targeted signals, stressing the necessity of rejecting same-half-plate pixel pairs. While the contamination is largely removed when we only use different-half-plate pixel pairs (left panel), there still appears to be a residual weak correlation at $\Delta\lambda_{\rm obs}\sim 0$, not expected from pure sky-subtraction effects. The exact source of such a weak correlation at $\Delta\lambda_{\rm obs}\sim 0$ may be related to some details in the processing procedure in the spectra pipeline. To proceed, we adopt a conservative method to remove the effect of this weak correlation by discarding any signal within $|r_\| |<4$cMpc, at the expense of slightly reducing the signal-to-noise ratio of our measurement.

\subsection{Large-scale Correction}

As discussed in \citet{Croft2016}, one may find non-zero cross-correlation for large pair separation with no physical significance we concern about. We correct this spurious signal by subtracting the average correlation over 80--400$\hinvMpc$ along both the line-of-sight and orthogonal directions, following the method described in \citet{Croft2016}. 

\section{Model for \lya\ Luminosity Density Contributed by Star-forming Galaxies}\label{sec: Lya model}

In our model, star-forming galaxies dominate the \lya\ luminosity density. We first review the \citet{Dijkstra2012} model for the REW distribution of \lya\ emission from an inner aperture around star-forming galaxies. With such an REW distribution, we present our model of \lya\ luminosity density from contributions of \lya\ emission within the inner aperture and from the outer halo.

\subsection{Model for Ly$\alpha$ Rest-frame Equivalent Width Distribution of UV-selected Galaxies}\label{sec:Dijkstra model}

\cite{Dijkstra2012} modelled the conditional probability density function (PDF) for the REW of \lya\ emission (from the central aperture around LBGs) using an exponential function whose scaling factor ${\REW}_c$ depends on $M_{\rm UV}$ and $z$,
\begin{equation}\label{eq: P_REW_Muv}
    P\left( {\REW }\mid M_{\mathrm{UV}}\right)= \begin{cases}\mathcal{N} \exp \left(-\frac{\mathrm{REW}}{\operatorname{REW}_{\mathrm{c}}}\right), & {\REW} \in\left(x_{\min }, x_{\max }\right) \\ 0, & \text { otherwise }\end{cases}
\end{equation}
where $\mathcal{N}$ denotes a normalization constant. The choice of the normalization factor $\mathcal{N}$ allows that all drop-out galaxies have $x_{\min}\leq{\REW}\leq x_{\max}$,
\begin{equation}
    \mathcal{N}^{-1}=\operatorname{REW}_{c} \left[\exp \left(-\frac{x_{\min }}{\operatorname{REW}_{\mathrm{c}}\left(M_{\mathrm{UV}}\right)}\right)-\exp \left(-\frac{x_{\max }}{\operatorname{REW}_{\mathrm{c}}\left(M_{\mathrm{UV}}\right)}\right)\right].
\end{equation}

To match the $M_{\rm UV}$-dependence of the observed fraction of LAEs  (${\REW}>50$\AA) in drop-out galaxies, they fixed $x_{\max}=300$ and assumed $x_{\min}\equiv -a_1$ (both in units of \AA),
\begin{equation}\label{eq:a1}
    a_{1}=\left\{\begin{array}{lcl}
    20 & & M_{\mathrm{UV}}<-21.5 \\
    20-6\left(M_{\mathrm{UV}}+21.5\right)^{2} & & -21.5 \leq M_{\mathrm{UV}}<-19 \\
    -17.5  &  &\text { other. }
    \end{array}\right.
\end{equation}

In their fiducial model, ${\REW_c}$ evolves with $M_{\rm UV}$ and $z$,  
\begin{equation}\label{eq:REW evolving model}
    {\REW}_c(M_{\rm UV},z) = {\REW}_{c,0} + \mu_1 (M_{\rm UV}+21.9) + \mu_2 (z-4),
\end{equation}
where the best-fitting parameters are ${\REW}_{c,0}=23$\AA, $\mu_1=7$\AA, $\mu_2=6$\AA. Note that the fitting formula applies only in the observed range of UV magnitudes and the evolution is frozen for $M_{\rm UV}>-19$.  However, in our analysis we adopt a constant ${\REW}_c=22$\AA, which depicts the REW distribution of the 400 brightest LBG sample of \cite{Shapley2003} well but underpredicts the faint-end LAE fraction, as disccussed in Appendix A1 of \cite{Dijkstra2012}. With this constant ${\REW}_c$, we would underestimate the \lya\ luminosity contributed by UV-faint galaxies, and the total estimated \lya\ emission would be a lower limit.

The \lya\ luminosity at a given $\REW$ and ${\UV}$ luminosity can be expressed as
\begin{equation}\label{eq: L_alpha_REW_Muv}
    L_\alpha\left({\REW},M_{\rm UV}\right) = L_{\rm UV,\nu} \left(\nu_{\alpha} / \lambda_{\alpha}\right)\left(\lambda_{\mathrm{UV}} / \lambda_{\alpha}\right)^{-\beta-2} \cdot {\REW}  ,
\end{equation}
with the absolute AB magnitude $M_{\rm UV} =-2.5 \log [{L_{\rm UV,\nu}}/({\rm erg\, s^{-1} Hz^{-1}})] +51.6$. The parameter $\beta$ characterizes the slope of the UV continuum, such that {$L_{\rm UV,\lambda}=\nu L_{\rm UV,\nu}/\lambda \propto \lambda^\beta$}. We adopt $\lambda_{\rm UV}=1700$\AA\ and fix $\beta=-1.7$ as in \citet{Dijkstra2012}. The adopted wavelength is the same as in the UV LF measurements (Table~\ref{tab:UV LF}), except for the \citet{Bouwens2015} UV LF (measured at 1600\AA). In our calculation, we ignore the slight wavelength shift in the \citet{Bouwens2015} UV LF, as the effect in the UV luminosity computation is less than 2\%.

\subsection{Model for the Inner and Outer Ly$\alpha$ Emission Component}

We separate star-forming galaxies into two populations based on the case of \lya\ radiation within the central 2$\arcsec$ aperture, one with \lya\ emission ($\REW>0$) and one with \lya\ absorption ($\REW<0$). We can express the corresponding UV LFs as
\begin{equation}\label{eq:phi_e}
\Phi^e_\UV(M_\UV)=\frac{\int_0^{+\infty} P(\REW\mid M_\UV) d\REW}{\int_{-\infty}^{+\infty}P(\REW\mid M_\UV) d\REW}\,\Phi_\UV(M_\UV)
\end{equation}
for the $\REW>0$ population and
\begin{equation}\label{eq:phi_a}
\Phi^a_\UV(M_\UV)=\frac{\int_{-\infty}^0 P(\REW\mid M_\UV) d\REW}{\int_{-\infty}^{+\infty}P(\REW\mid M_\UV)d\REW}\,\Phi_\UV(M_\UV)
\end{equation}
for the $\REW<0$ population, where $P(\REW\mid M_\UV)$ is the $\REW$ distribution for galaxies with UV luminosity $M_\UV$. Clearly, by construction, $\Phi^e_\UV+\Phi^a_\UV=\Phi_\UV$. Note that we formally use $-\infty$ and $+\infty$ for clarity, while the true cutoff thresholds are encoded in $P(\REW\mid M_\UV)$, which takes the form of Equation~(\ref{eq: P_REW_Muv}) if adopting the \citet{Dijkstra2012} model.

\begin{figure}
    \centering
    \includegraphics[width=\textwidth]{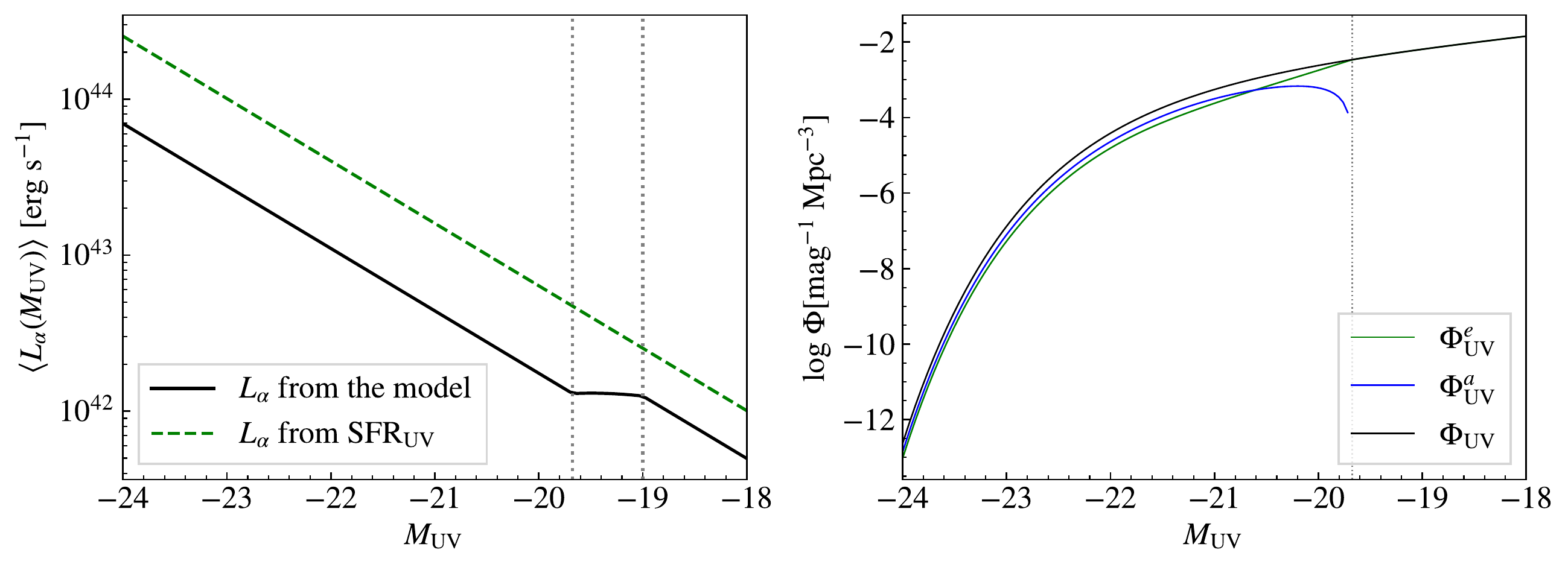}
    \caption{\textit{Left}: Mean \lya\ luminosity $\langle L_\alpha(M_\UV)\rangle$ within the 2$\arcsec$ aperture of the $\REW>0$\AA\ population as a function of the UV magnitude $M_\mathrm{UV}$, as presented in Equation~\ref{eq:mean_lya}. The gray dotted lines denote the turning points of $a_1$ as expressed in Equation~\ref{eq:a1}. The green dashed line denotes the expected \lya\ luminosity for the SFR associated with the UV luminosity. \textit{Right}: UV LF of the $\REW<0$\AA\ population $\Phi_{\rm UV}^a$, the $\REW>0$\AA\ population $\Phi_{\rm UV}^e$, and the entire population $\Phi_\UV$ as a function of $M_\UV$, as presented in Equation~\ref{eq:phi_e} and \ref{eq:phi_a}.  We take the \citet{Reddy2009} UV LF as an example. The gray dotted lines denotes one of the turning points of $a_1$ (Equation~\ref{eq:a1}), where $a_1=0$ and $\REW$ keeps larger than 0 as $M_\UV$ increases. That is, we assume that there is no $\REW<0$\AA\ population over this $M_\UV$ range, which will lead to an underestimation of the total \lya\ luminosity. 
    }
    \label{fig:Llya_and_Phi}
\end{figure}

The mean \lya\ luminosity within the 2$\arcsec$ aperture of the $\REW>0$ population at a given UV luminosity is
\begin{equation}\label{eq:mean_lya}
\langle L_\alpha(M_\UV)\rangle=\frac{\int_0^{+\infty}L_\alpha(\REW, M_\UV)P(\REW\mid M_\UV)d\REW}{\int_0^{+\infty} P(\REW\mid M_\UV)d\REW}.
\end{equation}
where $L_\alpha(\REW, M_\UV)$ can be calculated through Equation~(\ref{eq: L_alpha_REW_Muv}).  Figure~\ref{fig:Llya_and_Phi} presents the evolution of $\langle L_\alpha (M_\UV)\rangle$, $\Phi_\UV^a$ and $\Phi_\UV^e$ with $M_\UV$ in our model. We also show the expected \lya\ luminosity for the SFR associated with the UV luminosity, calculated through the relations that SFR of 1$M_{\odot}{\rm yr^{-1}}$ corresponds to UV luminosity $L_\nu=1.4\times 10^{-28} {\rm erg~s^{-1} Hz^{-1}}$ and \lya\ luminosity $L_\alpha = 1.1 \times 10^{42}{\rm erg~s^{-1}}$. It is much higher than our modelled \lya\ luminosity, consistent with the measurements in Figure~\ref{fig:SFRD compare}. 

\add{In addition, the net absorption from the $\REW<0$ population will also make a negative contribution. The `absorbed' luminosity  could be descibed as

\begin{equation}
\langle L_\alpha^{\rm Abs}(M_{\rm UV} )\rangle=\frac{\int_{-\infty}^{0}L_\alpha({\rm REW}, M_{\rm UV})P({\rm REW}\mid M_{\rm UV})d{\rm REW}}{\int_{-\infty}^{0} P({\rm REW}\mid M_{\rm UV})d{\rm REW}},
\end{equation}
which would yield a negative value.

}

The contribution to the \lya\ luminosity density from the inner part  comes from \add{the emission of the $\REW>0$ population and the absorption of the $\REW<0$ population, which is
\begin{equation}
\rho^{\rm inner}_{\rm Ly\alpha} = \int_{M_{\rm UV,min}}^{M_{\rm UV,max}} \left[ \langle L_\alpha(M_\UV)\rangle \Phi^e_\UV(M_\UV) 
+
 \langle L^{\rm Abs}_\alpha(M_{\rm UV})\rangle \Phi^a_{\rm UV}(M_{\rm UV})
\right] dM_{\rm UV}.
\end{equation}
In our model the negative absorption component is actually insignificant compared to the emission one, with the former being about 1--4\% of the latter depending on the adopted UV LF.
}

Based on the finding in \citet{Steidel2011}, we assume that the \lya\ luminosity in the diffuse halo component is the same as that from the central aperture in the $\REW>0$ population and that the diffuse component in the $\REW<0$ population takes the same value at a given UV luminosity.
Then the contribution from the outer part \lya\ emission of the $\REW>0$ population has the same expression as in the above equation, while that from the $\REW<0$ population is obtained by replacing $\Phi^e_\UV$ with $\Phi^a_\UV$. The total outer part contribution from \lya\ halos is then
\begin{equation}
\rho^{\rm outer}_{\rm Ly\alpha} = \int_{M_{\rm UV,min}}^{M_{\rm UV,max}} \langle L_\alpha(M_\UV)\rangle \Phi_\UV(M_\UV) dM_\UV.
\end{equation}
We adopt $M_{\rm UV,min}=-24$ and $M_{\rm UV,max}=-12$ in our calculation. 

\add{The outer part \lya\ emission can have contributions from satellite galaxies in high-mass halos \citep[e.g.,][]{Momose2016,Lake2015,Mitchell2021}, while the UV LF used to compute the inner part \lya\ emission should already include the satellite population. Therefore, in our model there is a possibility of double-counting the  contribution of \lya\ emission from the satellites. From halo modelling of LBG clustering, \citet{Cooray2006} find that the contribution from satellites to the UV LF is at a level of $\sim 10^{-3}$--$10^{-2}$ over a wide luminosity range and that it becomes even lower at the faint end ($M_\UV > -17$). A similar result is also obtained by \citet{Jose2013}. These empirical results suggest that the contribution of satellite galaxies to the total cosmic \lya\ luminosity density is negligible, and we simply ignore the effect induced by possibly double-counting satellites here.}


Note that our model is just a rough esitmate of the total \lya\ luminosity, with systematics arising from both the \lya\ REW PDF and UV LFs. For example, the \cite{Dijkstra2012} REW PDF may underpredict the number of large REW systems, leading to an underestimate of the total \lya\ luminosity. On the other hand, the modelled REW PDF may not describe the number of galaxies with net absorption very well. However, these uncertainties would not change our main claim significantly. Future observations of UV luminosity dependent \lya\ REW distribution and measurements of UV LFs are expected to improve the modelling.

\bibliography{sample63}{}
\bibliographystyle{aasjournal}

\end{document}